\documentclass[11pt,twoside,lineno]{article}

\usepackage[english]{babel}
\usepackage{indentfirst}
\usepackage[labelfont=bf]{caption}

\usepackage{comment}

\usepackage[letterpaper,top=2cm,bottom=2cm,left=3cm,right=3cm,marginparwidth=1.75cm]{geometry}

\usepackage{amsmath}
\usepackage{graphicx}
\usepackage[colorlinks=true, allcolors=blue]{hyperref}

\usepackage{float}
\usepackage{makecell}

\usepackage{ragged2e}

\usepackage{authblk}
\title{\textbf{Dynamic Manipulation of Multiphase Fluid in Microgravity Using Photoresponsive Surfactant}}
\author[a,$\#$]{Xichen Liang}
\author[b,$\#$]{Kseniia M. Karnaukh} 
\author[d]{Qixuan Cao}
\author[c]{Marielle Cooper}
\author[c]{Hao Xu}
\author[b]{Ian Maskiewicz}
\author[e]{Olivia Wander}
\author[b,*]{Javier Read de Alaniz}
\author[c,*]{Yangying Zhu}
\author[c,*]{Paolo Luzzatto-Fegiz}

\affil[a]{Department of Chemical Engineering, University of California at Santa Barbara, Santa Barbara, California 93106-5070, USA}
\affil[b]{Department of Chemistry, University of California at Santa Barbara, Santa Barbara, California 93106-5070, USA}
\affil[c]{Department of Mechanical Engineering, University of California at Santa Barbara, Santa Barbara, California 93106-5070, USA}
\affil[d]{Department of Physics, University of California at Santa Barbara, Santa Barbara, California 93106-5070, USA}
\affil[e]{Department of Materials, University of California at Santa Barbara, Santa Barbara, California 93106-5070, USA}

\affil[$\#$]{X.L.and K.K. contributed equally to this work.}
\affil[*]{To whom correspondence should be addressed. E-mail: jalaniz@ucsb.edu, yangying@ucsb.edu, pfegiz@ucsb.edu}

\begin{document}
\maketitle


\begin{abstract}
 Control of bubble motion is essential for improving efficiency and creating new functionalities in electrochemistry, heat transfer, and biomedical systems. 
Photoresponsive surfactants enable bubble manipulation by creating surface tension gradients, inducing a “photo-Marangoni” flow under illumination, without needing any engineered substrates, by leveraging a reversible switch in molecular conformation.
Although previous studies have demonstrated bubble manipulation using photo-responsive surfactants, a comprehensive understanding of how fluid behavior is affected by critical parameters, such as bubble size, illumination, photo-switching kinetics, concentration, and adsorption/desorption kinetics, remains elusive. Advances have been limited by the complex multiphysics processed involved, and by the fact that earth-bound experiments cannot study bubble photo-Marangoni dynamics without interference from bubble buoyancy and photo-thermal convection. 
We elucidate the factors enabling fast photo-Marangoni-driven bubble motion, by performing microgravity experiments, enabled by a bespoke photo-surfactant, complemented by a detailed modeling framework.
We identify an optimal bubble size for migration, since smaller and larger bubbles incur weaker photo-Marangoni stresses and larger drag, respectively. Surfactants that switch rapidly under illumination drive fast migration, provided their reverse switch (in darkness) is much slower, yet not negligible. 
These foundational results enable the synthesis of next-generation photo-surfactants and photo-Marangoni manipulation across multiphase fluid systems.
\end{abstract}

\newpage

The manipulation of bubbles is crucial in applications that involve multiphase processes \cite{xiong2021stimuli,shao2015near,liang2022manipulation}, such as electrochemical reactions \cite{park2023solutal,lu2014ultrahigh}, power generation, \cite{wen2018three}, thermal desalination \cite{toth2020modelling}, biomedicine \cite{shakeri2021gold}, and microreactors \cite{han2021overcoming}. For example, superaerophobic electrodes can promote bubble departure and enhance reaction efficiency \cite{li2021superwetting}. Promoting bubble detachment can also delay the critical heat flux during boiling, which is critical for power plant safety and electronics cooling \cite{hou2024electronic}. Existing strategies to manipulate bubble dynamics include passively engineering the wettability and texture of substrates to control their interaction with bubbles \cite{li2019drop, zhu2020spontaneous,xiao2021bioinspired,wang2019dynamics}, and actively exerting a force on bubbles using thermal \cite{wang2019thermally}, mechanical \cite{zhang2019bio, jiao2021situ}, magnetic \cite{guo2019magnetocontrollable,ben2022underwater}, and electric \cite{yan2018electrowetting} stimuli. These methods often require complex fabrication or high energy input. 

Alternatively, dynamic manipulation of bubbles using light-responsive substrates \cite{gao2019underwater, huang2021light} or surfactants offers real-time reconfiguration ability, and can potentially achieve fast response and high spatial resolution with low energy input. In particular, photoresponsive surfactants can be directly used to control bubbles without having to engineer any substrates because surfactants naturally accumulate at fluid-fluid interfaces. This type of surfactant changes its molecular conformation when illuminated with low-intensity light of appropriate wavelengths, which induces a local change in surface tension, and thereby to generate a Marangoni flow that drives bubble motion \cite{shang2003photoresponsive,zhao2024manipulation,florea2014photo,xiao2021photocontrolled}. Multiphase fluid manipulation has been performed using various photoresponsive surfactants, such as azobenzene trimethyl-ammonium bromide (AzoTAB), donor-acceptor Stenhouse adducts (DASA) and spiropyrans (SP) \cite{seshadri2021influence,galy2020self}. Recent studies have demonstrated dynamically reconfigurable emulsions \cite{reifarth2022dual, biswas2023light}, adjustable friction in hydrogels \cite{chau2024photoresponsive}, controlled particle deposition on surfaces \cite{anyfantakis2017evaporative}, and the movement of droplets both within and on the surface of an immiscible liquid \cite{xiao2018moving, liang2024dynamic, diguet2009photomanipulation}. While previous studies have shown the effectiveness of photoresponsive surfactants in manipulating multiphase fluid motion, a quantitative understanding of the underlying mechanisms, including the effects of surfactant photoswitching kinetics, adsorption/desorption parameters, illumination conditions, and bubble size, on the bubble migration velocity, remains elusive. This is mainly due to the complex reaction and transport processes involved, and to the experimental challenges in isolating the surfactant-induced photo-Marangoni force from other coexisting effects such as bubble buoyancy and natural convection. 

In this study, we present a unique microgravity experiment in conjunction with a numerical modeling framework to elucidate how various factors, including photoswitching kinetics, surfactant adsorption and desorption process, concentration, and bubble size, affect photo-Marangoni-induced bubble migration. The microgravity environment minimizes bubble buoyancy and natural convection driven by illumination and photo-thermal conversion, allowing observation of bubble migration driven solely by the Marangoni force. With MCH-para photosurfactant, which isomerizes upon application of blue light, we demonstrate the migration of air bubbles in methanol with a velocity of up to 7.35\,mm/s using only 37.5\,mW/cm\textsuperscript{2} of light intensity. Experimental results show increasing bubble migration velocity as the bubble radius reduces from 3\,mm to 1\,mm and as the light exposure duration increases from 2\,s to 10\,s.  
With the aid of our numerical model, we find that migration velocity eventually reduces for smaller bubbles, revealing an optimal bubble radius around 1\,mm. This optimum is explained by a scaling model, showing that small bubbles support weaker photo-Marangoni stresses, whereas larger bubbles incur a relatively higher drag. We further find that surfactants that `forward-switch' rapidly under illumination drive fast migration, provided their `reverse' switch (in darkness) is much slower than the forward one, while still having a non-negligible effect on the concentration at the bubble's surface. We further elucidate the effect of concentration and adsorption/desorption ratio for each photo-isomer, and explain the underlying physical and photo-chemical mechanisms. The results from this study enable the synthesis of next-generation photo-surfactants, as well as fast photo-Marangoni manipulation across multiphase fluid systems.

\subsection*{Working principle and characterization of photoresponsive surfactants} 

\begin{figure*}[t!]
\centering
\includegraphics[width=\linewidth]{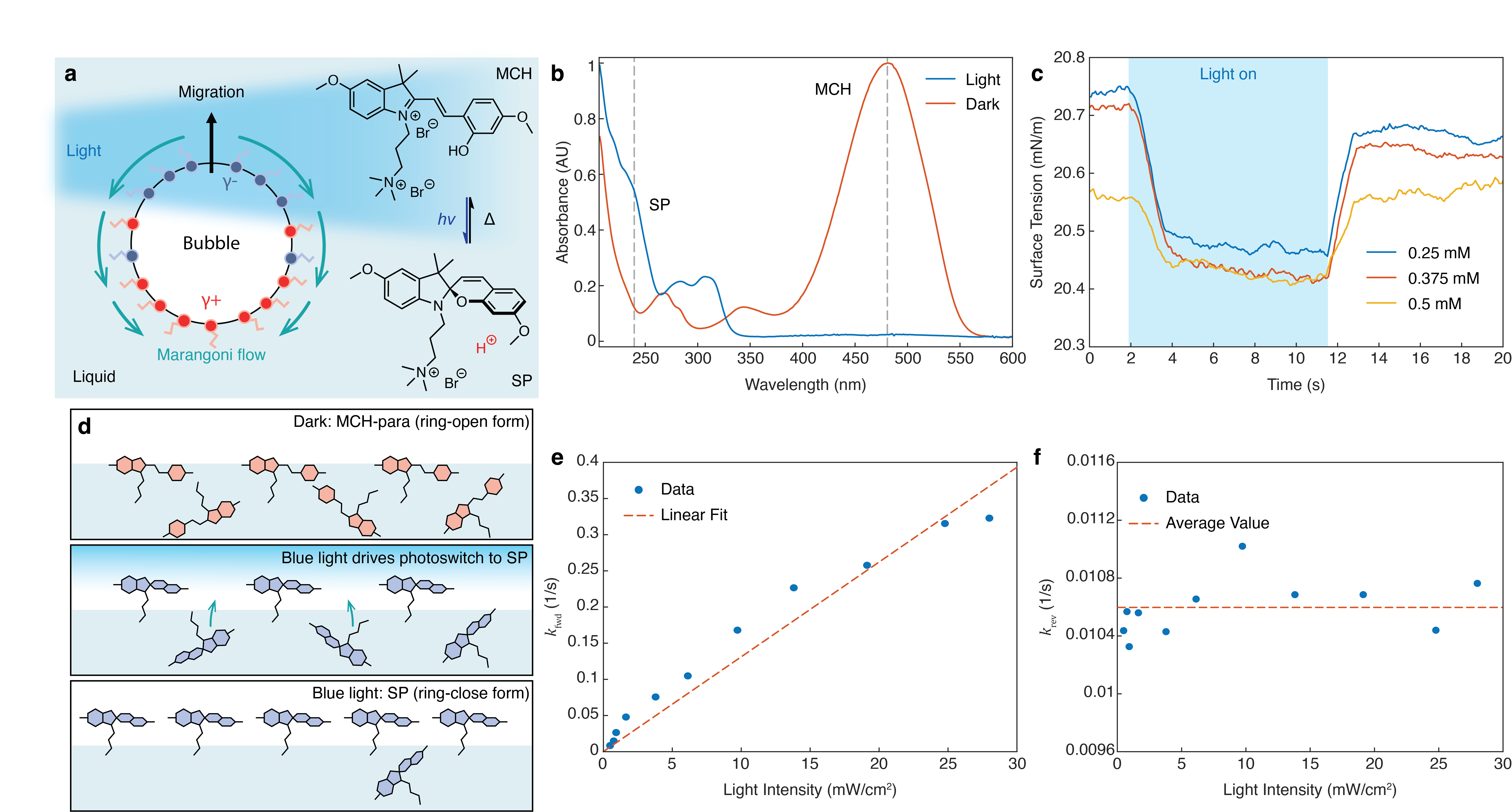}
\caption{\fontsize{9}{10}\selectfont
Mechanism of bubble migration driven by isomerization of photosurfactant MCH-para. 
\textbf{a.} Schematic of bubble migration in liquid, showing the molecular structures of the photoresponsive surfactant in ring-open (MCH) and ring-closed (SP) forms. Under illumination, the surface tension decreases, inducing a Marangoni flow from the illuminated region to the unilluminated region, which generates a net force toward the light. 
\textbf{b.} UV–Vis spectrum illustrating the switching mechanism of MCH-para in acidified methanol (0.025 mM). 
\textbf{c.} Surface tension response of MCH-para in methanol under 470 nm light (31.8 mW/cm\textsuperscript{2}) illumination at concentrations of 0.25 mM, 0.375 mM, and 0.5 mM.
\textbf{d.} Mechanisms of surface tension variation as MCH-para isomerizes to SP form under illumination. SP exhibits denser packing at the air-liquid interface, leading to a reduction in surface tension.
\textbf{e.} Forward reaction rate constants ($k_{\text{fwd}}$) and \textbf{f.} reverse reaction ($k_{\text{rev}}$) of MCH-para in acidified methanol (0.025 mM) as functions of light intensity.}
\label{fig:1} 
\end{figure*}

The control of bubble motion using the photo-Marangoni effect relies on creating surface shear stresses through surface tension gradients. As shown in Figure \ref{fig:1}a, when light is applied to a region of a bubble within a liquid medium, the illuminated surfactants undergo photoisomerization, changing their molecular conformation, and therefore their affinity for the fluid interface. An adsorptive flux leads to locally greater interfacial surfactant concentrations. This creates a gradient in surface tension that induces a Marangoni flow along the surface from the low-surface-tension side to the high-surface-tension side, propelling the bubble to migrate in the opposite direction. Dynamically adjusting the light allows for precise manipulation of the bubble without needing to engineer the solid substrates. Furthermore, many photosurfactants are sensitive to even very low-intensity light, making this approach attractive to applications prohibiting localized heating, as summarized in our previous work \cite{liang2024dynamic}.

The merocyanine-based photosurfactant, MCH-para (Figure \ref{fig:1}a), has fast photoswitching kinetics and excellent reversibility (Figure S6), capable of inducing rapid fluid flow upon blue light irradiation. It was previously synthesized and employed by us to dynamically manipulate droplet motion on liquid-infused surfaces \cite{liang2024dynamic}. The change in molecular conformation alters the UV-Vis ligt absorbance spectrum. Figure \ref{fig:1}b shows the UV–Vis spectrum illustrating the switching mechanism of MCH-para in acidified methanol (0.025 mM), where MCH-para isomerizes to spiropyran (SP) form  upon exposure to 470 nm light (SI Section 1.5). 

Surface tension (Figure \ref{fig:1}c) demonstrates a rapid response in methanol solutions of various concentrations under blue light exposure (470 nm, 31.8 mW/cm\textsuperscript{2}). The surface tension was measured using the pendant drop method on an optical tensiometer (SI Section 2.1). Figure \ref{fig:1}c shows that the surface tension response was very fast (within 2 seconds) across all concentrations. For this surfactant, surface tension decreases upon light illumination, and the magnitude of surface tension change decreases with increasing surfactant concentration. The decrease in surface tension can be qualitatively explained as illustrated in Figure \ref{fig:1}d. The change in surface tension is driven by the surface coverage of the two isomer states due to the balance of intermolecular forces at the air-liquid interface. Molecules in the more polar MCH form can orient themselves such that their polar regions interact with the methanol molecules. As illustrated in Figure \ref{fig:1}d, in the dark state, MCH-para exists in its less compact, ring-open form. This structure leads to the less dense packing of the surfactant at the air-liquid interface and allows the cohesive forces between the liquid molecules to remain relatively strong, maintaining a relatively high surface tension. Upon illumination, MCH-para switches to its ring-closed less polar isomer, SP, which has a more compact structure. This allows more SP molecules to occupy the surface, which effectively displaces methanol molecules and reduces the surface tension. 
Moreover, the concentration effect on the surface tension change in Figure \ref{fig:1}c is likely attributed to a higher concentration of MCH-para causing a greater surface coverage, resulting in lower surface tension in the initial state (i.e., before light illumination).   

In addition to the steady-state surface tension change, the rate of surface tension change is another important characteristic, which is related to the photoswitching kinetics of photosurfactants. We measured the forward and reverse reaction rate constants ($k_{\text{fwd}}$ and $k_{\text{rev}}$) for MCH-para isomerization using pump-probe spectroscopy under 11 different light intensities and fit the data with a pseudo-first-order reaction equation as expected for photo-isomerization (SI Section 1.6). Acidified methanol was used for 0.025 mM MCH-para solution preparation for UV-Vis and pump-probe spectroscopy experiments 
(SI Section 1.4). Examples of the fitted absorbance results at light intensities of 0.925 mW/cm\textsuperscript{2} and 28 mW/cm\textsuperscript{2} are shown in Figure S5. Figures \ref{fig:1}e and \ref{fig:1}f present the relationship between $k_{\text{fwd}}$ and $k_{\text{rev}}$ at different light intensities. The results show that $k_{\text{fwd}}$ increases approximately linearly with light intensity (Figure \ref{fig:1}e). Due to the maximum allowable optical intensity of the fiber-coupled LED, the $k_{\text{fwd}}$ at the microgravity experimental condition (37.5 mW/cm\textsuperscript{2}) was estimated by extrapolating a linear fit, yielding a value of 0.4875 s\textsuperscript{-1}. In contrast, $k_{\text{rev}}$ remained stable, averaging 0.0106 s\textsuperscript{-1} (Figure \ref{fig:1}f) across all light intensities. Moreover, the pump-probe spectroscopy demonstrates the rapid transitions of the MCH-para in methanol solution, as well as, good reversibility over multiple cycles (Figure S6). After repeated exposures, the photochemical transitions remain consistent within 5 cycles, suggesting that MCH-para can endure prolonged light irradiation. Due to the fast reproducible photoswitching kinetics, MCH-para was chosen for microgravity experiments on the International Space Station (ISS).

\subsection*{Microgravity experiment} 

\begin{figure*}[t!]
\centering
\includegraphics[width=\linewidth]{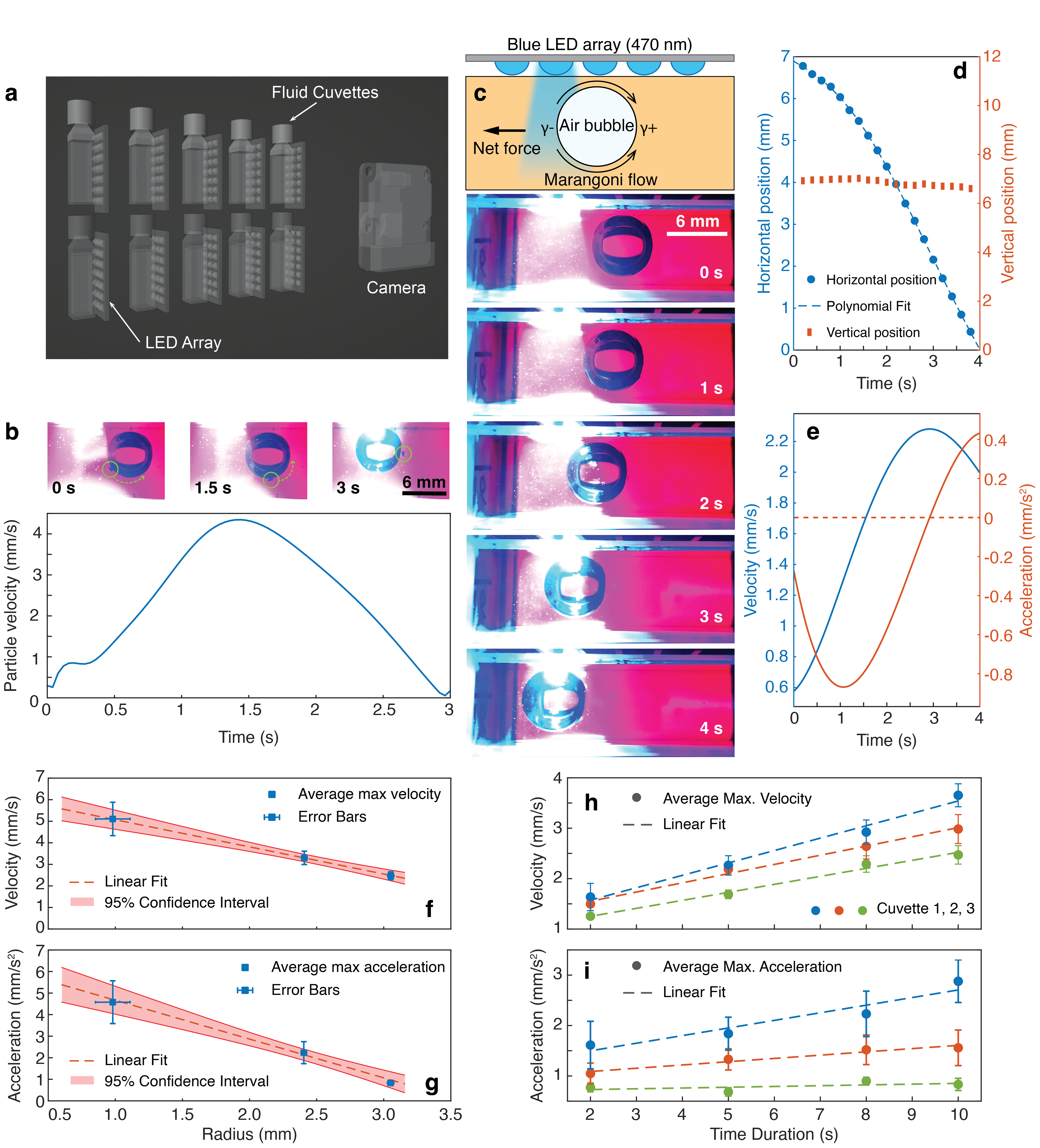}
\caption{\fontsize{9}{10}\selectfont
Experimental analysis of bubble migration under light exposure in microgravity. \textbf{a.} Schematic diagram of the experimental setup and its main components.
\textbf{b.} Velocity of a tracer particle over time during a single data point (from one light-on to light-off cycle) of exposure, illustrating velocity changes as the bubble transitions from partial to full illumination. Insets show images of the bubble at selected time points.
\textbf{c.} Time-lapse optical images showing bubble migration in microgravity, directed by a blue LED array.
\textbf{d.} Horizontal and vertical displacement of the bubble over time.
\textbf{e.} Line plot showing the velocity and acceleration of the bubble corresponding to panel \textbf{d}.
\textbf{f.} Average maximum migration velocity and \textbf{g.} acceleration as functions of bubble radius under 10 seconds of light exposure.
\textbf{h.} Average bubble migration velocity and \textbf{i.} acceleration as functions of light exposure duration.
}
\label{fig:2}
\end{figure*}

In conventional experiment, experimental observation of photo-Marangoni-driven bubble motion is non-trivial, as other effects, such as buoyancy, density-gradient-driven natural convection, and thermal-Marangoni force, are usually present. The coexistence of buoyancy can interfere with the study of the photo-Marangoni effect. On Earth, buoyancy dominates for bubbles with a radius larger than the capillary length (\textit{L} = 1.65 mm for methanol). A bubble in methanol with that radius can experience a buoyancy force on the order of 10\textsuperscript{-4} N, while the Marangoni force is approximately 10\textsuperscript{-7} N (SI Section 3). To eliminate buoyancy and natural convection, the experiments described in this paper were conducted onboard the International Space Station (ISS), which created a microgravity environment, as mentioned earlier. Although a thermal-Marangoni force caused by a surface tension gradient under a temperature gradient can still be present due to a mild heating effect from light absorption, we have carefully estimated its effect to be less than 11.6 percent (SI Section 4). 

The experimental setup consists of four-sided clear quartz cuvettes containing test fluids (1 mM MCH-para in methanol) and air bubbles of varying sizes (radius 1-3 mm), LED arrays, background lighting, high-resolution cameras, and temperature sensors fully enclosed in a CubeLab, as illustrated in Figure \ref{fig:2}a. The temperature and humidity inside the CubeLab for the duration of the experiment were approximately 29.31 °C and 16.62 \%H, respectively, which were monitored using a resistance temperature detector (RTD) and a PCB-mounted humidity sensor (SI Table 4). On the side of each cuvette, an array of low-power blue LED diodes (470 nm, optical intensity of 37.5 mW/cm\textsuperscript{2}) was arranged in a 3$\times$7 matrix to provide temporally- and spatially-controlled illumination. Each row of the LEDs was programmed to illuminate for a specific duration \textit{t} before the next row of LEDs was turned on for the same duration, and the experiments were repeated three times. Duration time \textit{t} was gradually increased from 2, 5, 8, to 10 seconds. Eight high-resolution cameras (Raspberry Pi V2.1, 25 fps, 1640 $\times$ 1232) were used to record the position and movement of bubbles in each cuvette.

The generated surface Marangoni flow was visualized using tracer particles, allowing for the estimation of fluid velocity near the bubble surface. Figure \ref{fig:2}b shows time-lapse images of a tracer particle flowing along the surface of a bubble 3 mm in radius; the particle velocity is also shown. To understand relative motion between particles and the bubble, the variations in particle velocity, reflecting dynamic changes as the bubble transitions between illuminated and dark regions, were adjusted to a steady frame of reference moving with the bubble. The velocity first increases as the bubble enters the light zone, where a higher concentration of switched SP molecules on the left side establishes a surface tension gradient. Following this peak, the velocity decreases as the bubble continues moving until it fully enters the light zone, whereby the surface tension gradient diminishes due to the more uniform distribution of the isomers.

Figure \ref{fig:2}c shows the side-view schematic of one cuvette containing the test solution, the bubble, illuminated by the LED array, and a series of representative, recorded time-lapse optical images (See SI Video S1) of a bubble 3 mm in radius exposed to light for 10 s. When the bubble is near the illuminated zone, the photo-Marangoni effect generates a shear force that propels it toward the illuminated zone (0 s to 3 s). As the bubble becomes fully illuminated, the surface tension gradient diminishes, leading to a reduction in migration velocity (3 s to 4 s). Each time the LED light is turned on, the bubble is attracted from the dark region to the illuminated zone, constituting a segment. The time-dependent position of the bubble was tracked using MATLAB (Figure \ref{fig:2}d), which was further analyzed to obtain the migration velocity and acceleration (Figure \ref{fig:2}e). In this example, the maximum acceleration reached 0.78 mm/s\textsuperscript{2}, with a peak velocity of 2.26 mm/s. Across all experiments, the highest recorded migration velocity in methanol was 7.35 mm/s. Videos of more experimental conditions are presented in the SI Section 5 and SI Video S1.

\subsection*{Visualization of fluid motion: effects of bubble size and light exposure duration} 

The bubble migration velocity and acceleration were observed to be dependent on the size of the bubble and on the duration of light exposure. Figure \ref{fig:2}f and \ref{fig:2}g show the average of maximum bubble migration velocity and the average of maximum acceleration under 10 seconds of light exposure, categorized into three groups with bubble sizes of 1 mm, 2.5 mm, and 3 mm. The results suggest that both the velocity and acceleration increase as bubbles decrease in size from 3 mm to 1 mm. Across the experiments, the average maximum velocity and acceleration for bubbles of 1 mm are 5.1 mm/s and 4.6 mm/s\textsuperscript{2}, respectively. 

Figure \ref{fig:2}h and \ref{fig:2}i illustrate the effects of light exposure duration on bubble migration velocity and acceleration. Three cuvettes containing the same test fluid (1 mM MCH-para in methanol) were studied, with light exposure durations controlled at 2 s, 5 s, 8 s, and 10 s. In each cuvette, both the migration velocity (Figure \ref{fig:2}h) and acceleration (Figure \ref{fig:2}i) of bubbles increase with increasing light exposure duration. This trend suggests that longer illumination enhances the Marangoni force acting on the bubble by allowing more MCH-para to undergo isomerization. With a $k_{\text{fwd}}$ of 0.4875 s\textsuperscript{-1}, the fraction of molecules switched increases from 0.62 at 2 s to 0.97 at 10 s (SI Section 2.2). Additionally, prolonged illumination contributes to accumulated heat in the illuminated zone (estimated 11.6\% contribution) (SI Section 4). The combined effects of isomerization and localized heating further extend the surface tension gradient, leading to stronger bubble propulsion.

\subsection*{Modeling framework and finite element simulations}

\begin{figure*}[t!]
\centering
\includegraphics[width=\linewidth]{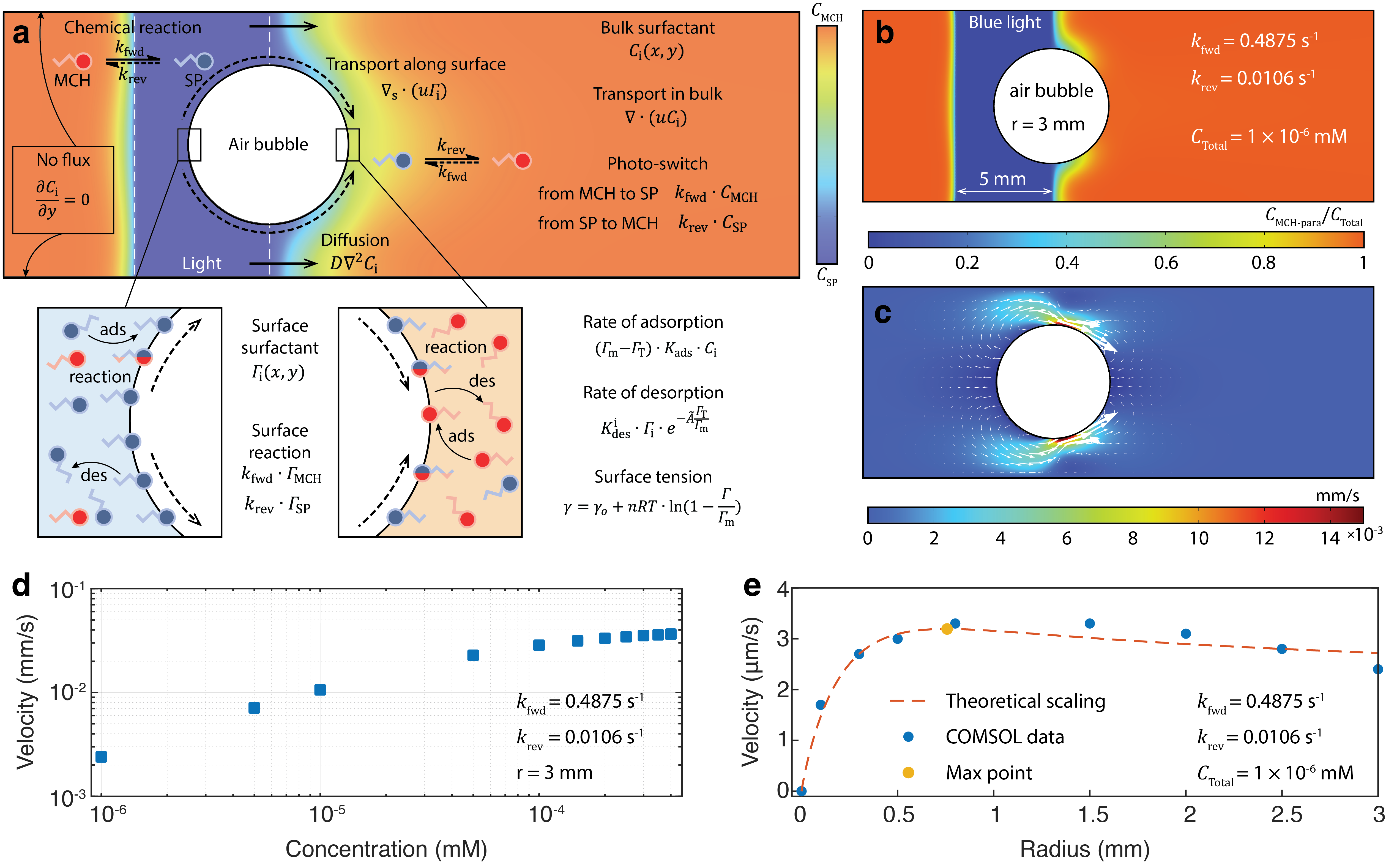}
\caption{\fontsize{9}{10}\selectfont
COMSOL simulations and theoretical analysis illustrate the influence of photoresponsive surfactants on bubble migration under light exposure. \textbf{a.} Schematic representation of the physics involved in bubble migration under light exposure in the presence of photoresponsive surfactants. The diagram illustrates the transport processes in both the bulk solution and along the surface, including diffusion, adsorption, desorption, and the reversible photoswitching reaction between two forms, MCH and SP. The rate of adsorption, desorption, and surface reaction is depicted, along with their influence on surface tension. The color gradient reflects the concentration distribution of the surfactants.
\textbf{b.} COMSOL simulations illustrations of MCH-para concentration distribution and \textbf{c.} fluid velocity field around a bubble with a 3 mm radius, with \(k_{\text{fwd}}\) = \(0.4875 \, \mathrm{s^{-1}}\) and \(k_{\text{rev}}\) = \(0.0106 \, \mathrm{s^{-1}}\). 
\textbf{d.} Simulated bubble migration velocity as a function of surfactant concentration for a 3 mm radius bubble,  with \(k_{\text{fwd}}\) = \(0.4875 \, \mathrm{s^{-1}}\) and \(k_{\text{rev}}\) = \(0.0106 \, \mathrm{s^{-1}}\). 
\textbf{e.} Bubble migration velocity as a function of bubble radius, with dashed lines representing theoretical velocities at small (blue) and large (orange) Reynolds numbers (Re), and a combined model (yellow) fitting the simulation data points (purple). The total concentration is controlled at 10\textsuperscript{-6} mM, with \(k_{\text{fwd}}\) = \(0.4875 \, \mathrm{s^{-1}}\) and \(k_{\text{rev}}\) = \(0.0106 \, \mathrm{s^{-1}}\).}
\label{fig:3} 
\end{figure*}

A modeling framework was developed to investigate the effects of surfactant switching kinetics, diffusion and adsorption coefficients, and bubble size on bubble migration velocity. The insights gained from the model provide guidelines for the design of photoresponsive surfactants for precise multiphase fluid manipulation. This model integrates advection and chemical reactions of the surfactants both in bulk and along the surface, as well as adsorption/desorption at the interface (Figure \ref{fig:3}a) to predict fluid velocity field, the concentration distribution between two isomers, and the bubble migration velocity. A complete set of equations in the modeling framework is presented in the SI Section 6. The model is constructed with the left half of the bubble illuminated while the rest of the domain remains in darkness. In the illuminated region, molecules primarily isomerize from the MCH form to the SP. In the dark zone, only the reverse thermal relaxation reaction from SP to MCH occurs. A Marangoni stress forms on the bubble surface as a result of the local concentration and species of adsorbed surfactants. The reference frames in both the experimental setup and the modeling framework are illustrated in Figure S10. In the space experiment, a bubble is inside a sealed cuvette filled with fluid, with stationary walls and the bubble moving towards the left. To make the problem tractable, we assume that the model is at steady state when observed in the reference frame moving with the bubble. To support the steady-state assumption, inflow/outflow boundary conditions are applied at the cuvette ends. In other words, the reference frame follows the bubble, and the velocity solved from the model is relative to the bubble migration velocity. The bubble's migration velocity is, therefore, equal to the bulk fluid velocity at the inlet. The model is solved numerically in COMSOL Multiphysics. 

Figure \ref{fig:3}b and \ref{fig:3}c illustrate an example of the steady-state concentration distribution of MCH-para and fluid velocity field around a bubble with a radius of \(3 \, \mathrm{mm}\) in a \(10^{-6} \, \mathrm{mM}\) concentration solution. The forward and reverse rate constants, \(k_{\text{fwd}}\) and \(k_{\text{rev}}\), are set to \(0.4875 \, \mathrm{s^{-1}}\) and \(0.0106 \, \mathrm{s^{-1}}\), respectively, which are determined experimentally from Figure \ref{fig:1}e and \ref{fig:1}f. In Figure \ref{fig:3}b, only a low concentration of MCH isomer remains in the light-exposed zone which indicates a high concentration of photoswitched SP due to the high \(k_{\text{fwd}}\) and low \(k_{\text{rev}}\). The newly formed SP molecules are transported along the bubble surface by Marangoni flow and diffusion. This distribution of MCH and SP molecules generates shear stress on the bubble surface, particularly from the point where the SP molecules exit the light zone and move to the rear side of the bubble, resulting in a high-velocity region (Figure \ref{fig:3}c).

To investigate the effect of surfactant concentration on bubble migration velocity, Figure \ref{fig:3}d further shows that the bubble migration velocity increases as the surfactant concentration increases from 10\textsuperscript{-6} to 10\textsuperscript{-3} mM, with the bubble size maintained at a radius of 3 mm and the rate constants \(k_{\text{fwd}}\) and \(k_{\text{rev}}\) set to \(0.4875 \, \mathrm{s^{-1}}\) and \(0.0106 \, \mathrm{s^{-1}}\), respectively. However, the rate of increase in velocity tapers off at higher concentrations. This occurs because, at higher concentrations, the surface coverage in the illuminated reion approaches saturation, limiting the adsorption of additional molecules from the bulk solution. To ensure sufficient photoresponsive surfactants during the space experiment, even in the event that a portion of the photo-surfactant were to degrade due to unforeseen issues, the concentration of MCH-para in methanol solution was 1 mM in the ISS experiment. When simulating higher concentrations, sharp gradients in surface concentration developed near the bubble stagnation points, which required significantly refined numerical meshes to ensure convergence, leading to large calculation times. For this reason, here we limit the range of concentrations from 10\textsuperscript{-6} to 10\textsuperscript{-3} mM.

Figure \ref{fig:3}e shows the bubble migration velocity for varying radii up to 3 mm at a fixed concentration of $10^{-6}$ mM. The maximum velocity occurs at a bubble radius of approximately 1 mm (yellow marker). The numerical finding that migration velocity decreases as r increases past r $\approx$ 1 mm is consistent with our ISS experiments, shown earlier in Figure \ref{fig:2}f. However, our simulations reveal that the velocity also decreases for bubbles smaller than r $\approx$ 1 mm. To explain this trend, a theoretical scaling was derived (SI Section 7), and is plotted alongside the numerical simulations, represented by the dashed orange line in \ref{fig:3}e. The trends can be understood as follows. For small bubbles (r $\leq$ 1 mm), the distance for a surfactant molecule to travel from the illuminated zone to the dark zone is short, which results in an incomplete reverse reaction and a weak surface tension gradient or Marangoni flow.  For large bubbles (r $\geq$ 1 mm), the migration velocity ultimately decreases as the drag force increases slowly with bubble size.

\begin{figure*}[htbp]
\centering
\includegraphics[width=\linewidth]{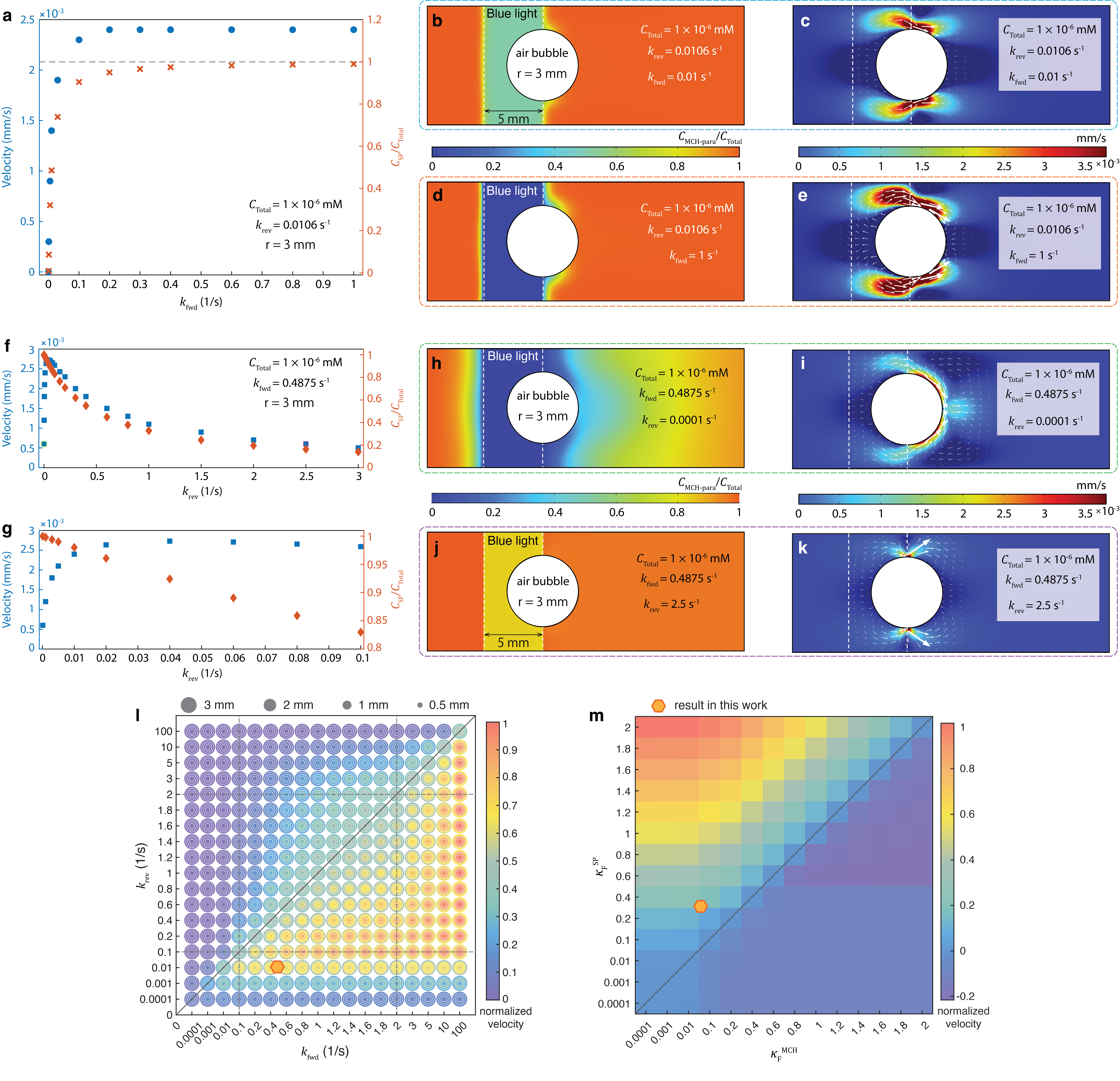}
\caption{\fontsize{9}{10}\selectfont
 Effect of reaction kinetics and adsorption/desorption rate ratios on bubble migration velocity.
\textbf{a.} Bubble migration velocity and fractional concentration of SP $C_{\text{SP}}/C_{\text{Total}}$ in the illuminated region as a function of $k_{\text{fwd}}$.  $k_{\text{rev}} = 0.0106$ s\textsuperscript{-1}. $C_{\text{Total}} = 1 \times 10^{-6}$ mM, and the bubble radius is 3 mm.
\textbf{b.} Concentration distribution of MCH around an air bubble (radius = 3 mm) with $k_{\text{fwd}} = 0.01 \, \text{s}^{-1}$. The left half of the bubble is illuminated. $C_{\text{Total}} = 1 \times 10^{-6}$ mM and $k_{\text{rev}} = 0.0106 \, \text{s}^{-1}$.
\textbf{c.} The velocity profile around the bubble for the same conditions as in \textbf{b}.
\textbf{d.} Concentration distribution of MCH with $k_{\text{fwd}} = 1 \, \text{s}^{-1}$ under the same conditions as in \textbf{b}.
\textbf{e.} The velocity profile around the bubble for the same conditions as in \textbf{d}.
\textbf{f.} Effect of varying $k_{\text{rev}}$ on bubble migration velocity and fractional concentration of SP $C_{\text{SP}}/C_{\text{Total}}$ in the illuminated region. $k_{\text{fwd}} = 0.4875$ s\textsuperscript{-1}. $C_{\text{Total}} = 1 \times 10^{-6}$ mM, and the bubble radius is 3 mm.
\textbf{g.} Zoomed-in view of \textbf{f} for $k_{\text{rev}}$ in the range 0 to 0.1 s\textsuperscript{-1}.
\textbf{h.} Concentration distribution of MCH around a 3 mm air bubble in the illuminated region with $k_{\text{rev}} = 0.0001$ s\textsuperscript{-1}, $k_{\text{fwd}} = 0.4875$ s\textsuperscript{-1}, and $C_{\text{Total}} = 1 \times 10^{-6}$ mM.
\textbf{i.} The velocity profile around the bubble for the same conditions as in \textbf{h}.
\textbf{j.} Concentration distribution of MCH around a 3 mm bubble with $k_{\text{rev}} = 2.5$ s\textsuperscript{-1}.
\textbf{k.} The velocity profile for the same conditions as in \textbf{j}.
\textbf{l.} Normalized bubble migration velocity as a function of reaction rate constants and bubble size. Color in each circle indicates migration velocity for different bubble radii. The solid gray line represents equal forward and reverse reaction rates, while the dashed gray line highlights the logarithmic and semi-logarithmic scales.
\textbf{m.} Influence of the Frumkin constants, $\kappa_F^{\text{MCH}}$ and $\kappa_F^{\text{SP}}$, on bubble migration velocity, representing the adsorption ratios of MCH and SP molecules 
The color scale shows normalized migration velocity. Negative values indicate reverse motion. The diagonal line represents equal adsorption rates for MCH and SP, producing a zero velocity, and the orange hexagon marks the experimental result. 
}
\label{fig:4} 
\end{figure*}

The model further enables us to investigate how the reaction kinetics and adsorption/desorption properties of the photosurfactants affect bubble migration, which can inform the optimal design of photosurfactants for dynamic fluid manipulation. 
To investigate how the forward reaction constant $k_{\text{fwd}}$ and reverse reaction constant $ k_{\text{rev}}$ each affect bubble migration, we first vary $k_{\text{fwd}}$ while keeping $k_{\text{rev}}$ constant at 0.0106 s\textsuperscript{-1}. As shown in Figure \ref{fig:4}a, as $k_{\text{fwd}}$ increases from 0 to 1 s\textsuperscript{-1} (such that $k_{\text{fwd}}/k_{\text{rev}}$ increases from 0 to 100), both the concentration of SP in the illuminated region and the bubble migration velocity increase sharply and then saturate after $k_{\text{fwd}}$ reaches 0.2 s\textsuperscript{-1}. This is because, for small $k_{\text{fwd}}$, MCH-para switches to SP slowly, resulting in a high concentration of MCH-para across the entire surface (Figure \ref{fig:4}b). This produces a low surface tension gradient and consequently slow motion (Figure \ref{fig:4}c). However, if the forward reaction is very fast, corresponding to high $k_{\text{fwd}}$, nearly all MCH-para molecules exposed to light are switched to SP, as indicated by the low concentration of MCH-para shown in Figure \ref{fig:4}d. Under this condition, since the forward reaction has effectively saturated, the surface tension gradient is mainly affected by the reverse reaction on the dark side of the bubble, as shown in Figure \ref{fig:4}e. Further increasing the forward reaction rate does not contribute to a faster bubble motion.

We then explore how variations in the reverse reaction constant $k_{\text{rev}}$ affect the bubble migration velocity. For this purpose, $k_{\text{fwd}}$ is fixed at 0.4875 s\textsuperscript{-1}. Surprisingly, as shown in Figure \ref{fig:4}f, when $k_{\text{rev}}$ increases from 0 to 3 s\textsuperscript{-1}, the migration velocity increases first and then decreases. Figure \ref{fig:4}g provides a zoomed-in view for $k_{\text{rev}}$ between 0 and 0.1 s\textsuperscript{-1}. As $k_{\text{rev}}$ increases from 0 to approximately 0.05 s\textsuperscript{-1}, more SP molecules will switch back to MCH-para in the dark half of the bubble, effectively increasing the surface tension gradient. This explains the initial increase in the bubble migration velocity. However, as $k_{\text{rev}}$ increases further, more of the switched SP will reverse back to MCH-para while still in the illuminated zone, where both the forward and reverse reactions take place. This is illustrated by the low concentration of SP in the illuminated zone as shown in Figure \ref{fig:4}j. In other words, if the reverse reaction is significantly faster than the forward reaction ($k_{\text{rev}}$ $\gg$ $k_{\text{fwd}}$), the bubble surface will be rich in MCH-para, resulting in a small surface tension gradient and low bubble migration velocity. The effect of $k_{\text{rev}}$ on bubble migration is further illustrated in Figure \ref{fig:4}h-k, where the extremes of $k_{\text{rev}}$ (0.001 s\textsuperscript{-1} and 2.5 s\textsuperscript{-1}) were selected. Both values of $k_{\text{rev}}$ lead to the same migration velocity of only $v = 6 \times 10^{-4}$ mm/s. If the reverse reaction is too slow ($k_{\text{rev}}$ = 0.001 s\textsuperscript{-1}, Figure \ref{fig:4}h), the entire bubble is surrounded by the switched SP molecules. On the other hand, if the reverse reaction is too fast ($k_{\text{rev}}$ = 2.5 s\textsuperscript{-1}, Figure \ref{fig:4}j), the bubble is surrounded by MCH-para. Both cases lead to low surface tension gradients.  The corresponding velocity distributions are illustrated in Figure \ref{fig:4}i and k, where a slow reverse reaction leads to a gradual change of surface tension in the dark zone (Figure \ref{fig:4}i), and a fast reverse reaction lead to a sudden change of surface tension localized at the boundary of the light beam (Figure \ref{fig:4}k).

Based on the above analysis, the desired conditions for fast bubble migration are: (1) a fast forward reaction to produce a nearly complete switch of MCH-para to SP in the illuminated zone, and (2) a moderately fast reverse reaction rate such that the reverse reaction occurs gradually over the entire length of the dark half of the bubble, and is not too fast to switch SP back to MCH-para in the illuminated zone. It is clear from the second condition that the optimal reverse reaction rate is affected by the size of the bubble.  Figure \ref{fig:4}l shows the normalized bubble migration velocity under different combinations of $k_{\text{fwd}}$, $k_{\text{rev}}$, and bubble size at a concentration of 10\textsuperscript{-6} mM. At any given $k_{\text{fwd}}$ and $k_{\text{rev}}$, the velocities of bubbles with radii of 0.5 mm, 1 mm, 2 mm, and 3 mm are represented by concentric circles with increasing radius. The color scale represents the normalized velocity relative to the maximum velocity in this parametric sweep plot. The orange hexagon marks the reaction kinetics of the photosurfactant used in the space experiment. Consistent with our previous analysis, the fastest bubble motion is achieved with the highest $k_{\text{fwd}}$ for bubbles with radii of 0.5 mm and 1 mm. It is worth noting that the faster the forward reaction $k_{\text{fwd}}$, the less sensitive the optimal migration velocity is to the choice of $k_{\text{rev}}$. For example, at $k_{\text{fwd}}$=100 s\textsuperscript{-1}, the optimal velocity can be achieved when $k_{\text{rev}}$ varies two orders of magnitude (from 0.1 to 10 s\textsuperscript{-1}). In comparison, when $k_{\text{fwd}}$=1 s\textsuperscript{-1}, the optimal velocity condition is only met when $k_{\text{rev}}$ is between 0.1 and 0.2 s\textsuperscript{-1}. This is because if the forward reaction is extremely fast, $k_{\text{rev}}$ needs to be extremely high to revert SP back to MCH-para in the illuminated zone. The lower bound of $k_{\text{rev}}$ for the optimal migration velocity across a wide range of $k_{\text{fwd}}$ remains relatively constant at approximately 0.1 s\textsuperscript{-1}. This is consistent with the criterion that SP molecules need to switch back to MCH-para when they travel to the rear end of the bubble in the dark zone. Any lower reverse reaction rate will lead to an incomplete switching which reduces the surface tension gradient.

We use our model to also investigate the influence of the adsorption and desorption rates of surfactants on the migration velocity. Figure \ref{fig:4}m shows how the Frumkin constants, $\kappa_F^{\text{MCH}}$ and $\kappa_F^{\text{SP}}$, which describe the ratio of the adsorption rate and the desorption rate of MCH-para and SP molecules, affect migration velocity (further details on these simulations are in SI section 2.2). The color scale represents the normalized migration velocity relative to the maximum velocity in this set of simulations. Negative values indicate reverse motion. The diagonal line represents equal adsorption rates for both MCH and SP molecules, while the orange hexagon marks the experimental condition in this study. In the upper-left region, a higher $\kappa_F^{\mathrm{SP}}$	leads to greater surface coverage of the switched SP molecules on the illuminated side. This lowers the local surface tension, creating a strong surface tension gradient that drives fast bubble motion toward the light source. Along the diagonal line, where $\kappa_F^{\mathrm{SP}}$ and $\kappa_F^{\mathrm{MCH}}$ result in equal surface coverage, no surface tension gradient is established, and thus no motion occurs. In the lower-right region, where $\kappa_F^{\mathrm{MCH}}$ is higher than $\kappa_F^{\mathrm{SP}}$, MCH molecules dominate the bubble surface in the dark region. Meanwhile, the switched SP molecules in the illuminated region exhibit reduced adsorption ability, leading to lower surface coverage in that zone. Consequently, the surface tension is lower on the MCH-rich side in the dark, generating a reversed surface tension gradient that propels the bubble in the opposite direction.


\section*{Conclusion}
In this work, we demonstrate dynamic and programmable control of bubble migration using photosurfactant MCH-para in microgravity, which isolates the Marangoni effect from natural convection and buoyancy. Experimental results suggest that the bubble migration velocity has a strong dependence on bubble size and the duration of light exposure. A modeling framework was constructed and solved numerically, which interrogates the influence of the bubble size, surfactant concentration, adsorption/desorption rate, and the switching kinetics of the surfactants on bubble migration velocity. The simulation identifies an optimal bubble radius of approximately 1 mm for maximum migration velocity with MCH-para, due to the competition of surfactant switching rate and fluid drag. Furthermore, a parametric study was conducted to determine the optimal switching kinetics and the adsorption/desorption rates of the surfactants for fast bubble migration. A combination of fast forward switching rate, intermediate reverse switching rate, and high adsorption rate of the switched isomeric molecules contribute to a high surfactant concentration gradient and thus lead to a high bubble migration velocity. The findings serve as important design guidelines for photoresponsive surfactants and their application in dynamic multi-phase fluid control. The modeling framework can be further extended to study liquid-liquid systems and their interactions with solid surfaces. The insights offer exciting opportunities for multi-phase systems in thermal, biomedical, catalytic, energy, and water applications both on earth and in space.

\section*{Materials and Methods}

\subsection*{Construction of CubeLab}
The ISS National Laboratory offers a long-term microgravity environment that is crucial for eliminating buoyancy forces on bubbles and natural convection.  A "CubeLab"-based setup was constructed following recommendations from CASIS. CubeLabs are self-contained experimental units that can be operated remotely from ground stations.

\subsection*{Sample preparations}
The surfactant MCH-para (molecular weight: 584.393 g/mol) was synthesized in-house (see SI Section 1 for detailed protocol). Methanol, used as the solvent, was obtained from Sigma-Aldrich (322415-1L). Test fluids were prepared by enclosing 1 mM MCH-para in four-sided clear quartz cuvettes, containing air bubbles with radii ranging from 1 to 3 mm. These fluids demonstrated stability over a broad temperature range for more than a month. Experimental temperatures within the CubeLab were measured at approximately 30$^\circ$C.

\subsection*{Light sources}
The fiber-coupled 470 nm blue LEDs used for UV-Vis and surface tension measurements (M470F3, 17.2 mW minimum fiber-coupled LED, and M470L3-C1, 350 mW collimated LED) were purchased directly from Thorlabs, Inc. The blue LED light panel used in the CubeLab was obtained from Digi-Key (SST-10-SB-B90-M470) and provided light in the 460–480 nm wavelength range. A 3D-printed frame was used to shape the light beam and achieve the desired illumination pattern. The LEDs illuminated in sequences of three in a row, cycling on and off continuously. The illumination duration for each experiment was set at 2, 5, 8, or 10 seconds, with the light cycling from left to right and then back from right to left.

\subsection*{Instrument and Characterization}
Surface tension was characterized using a standard pendant drop and pendant bubble method on commercial tensiometer, Biolin Scientific, Theta Flex. Optical power was measured by using optical power meter (Newport, 843-R) connected with a thermopile Sensor (Newport 919P-003-10, 3 W, 10 mm, 0.19-11~$\mu$m). Videos were taken by high-resolution camera (Raspberry Pi V2.1, 25 fps, 1640 $\times$ 1232). UV–Vis absorption spectra were recorded on an Agilent 8453 UV–Vis spectrometer from 200 to 1200 nm wavelengths. \textsuperscript{1}H and \textsuperscript{13}C NMR spectra were recorded on a Bruker 500 MHz NMR spectrometer.

\subsection*{Data Availability}
The data described in this article are available in figshare at 10.6084/m9.figshare.28577618.

\section*{Acknowledgements}

This research is supported by NSF grant titled “ISS: Dynamic Manipulation of Multi-Phase Flow Using Light-Responsive Surfactants for Phase-Change Applications”, grant number 2025655, and sponsored by ISS National Lab with Dr. Ryan Reeves as the program manager and Shawn Stephens as the operations project manager. We would like to acknowledge Space Tango as our industry partner who provided support of the space experimental hardware.

\bibliographystyle{unsrt}
\bibliography{reference}

\newpage

\section*{Supporting Information}
\section{Synthesis and Characterization of MCH-para}
\subsection{Materials and Instrumentation}
Solvents and reagents were purchased from commercial vendors and used without further purification unless otherwise specified. Anhydrous methanol (MeOH), 99.8\%, was purchased from Sigma Aldrich. Acetonitrile (CH\textsubscript{3}CN) for reactions was dispensed from a solvent purification system. Ethanol (EtOH) was stored over 3Å molecular sieves. Analytical thin-layer chromatography (TLC) was carried out with Merck TLC plates (silica gel 60 F254 on aluminum) and visualized by exposure to UV light (254/366 nm). Silica gel chromatography was performed using silica gel (60 Å pore size, 40 – 63 $\mu$m particle size). \textsuperscript{1}H and \textsuperscript{13}C NMR spectra were recorded at 298 K on a Varian Unity Inova AS600 600 MHz or Bruker Avance NEO 500 MHz NMR spectrometer using CDCl\textsubscript{3}, DMSO-\textit{d}\textsubscript{6} or CD\textsubscript{3}OD as the solvent. Chemical shifts ($\delta$) are reported relative to residual solvent peaks ($\delta$ 7.26 ppm for CDCl\textsubscript{3}, 2.50 ppm for DMSO-\textit{d}\textsubscript{6}, 3.31 ppm for CD\textsubscript{3}OD in \textsuperscript{1}H NMR; $\delta$ 77.16 for CDCl\textsubscript{3}, 39.52 ppm for DMSO-\textit{d}\textsubscript{6}, 49.00 ppm for CD\textsubscript{3}OD in \textsuperscript{13}C NMR).\\

\subsection{Synthesis of MCH-para}
(\textit{E})-2-(2-hydroxy-4-methoxystyryl)-5-methoxy-3,3-dimethyl-1-(3-(trimethylammonio)propyl)-3\textit{H}-indol-1-ium

\begin{figure*}[h]
\centering
\includegraphics[width=\linewidth]{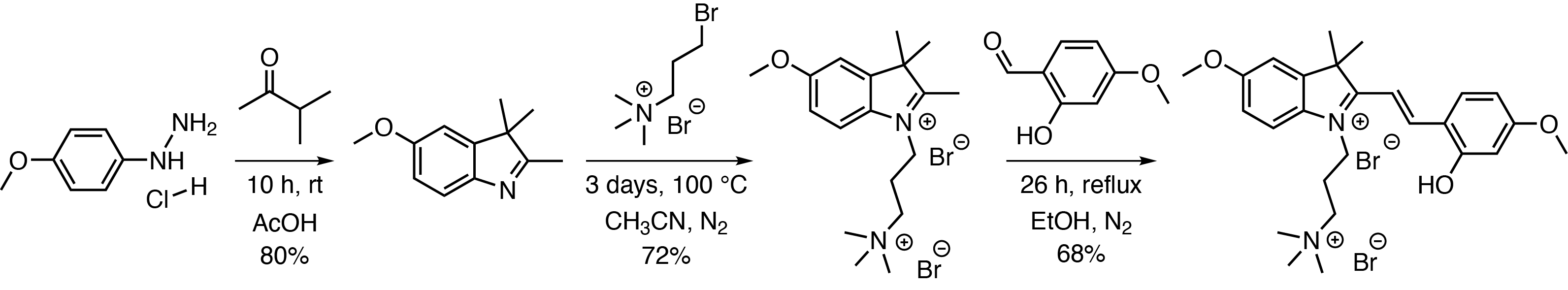}
\end{figure*}

\textbf{MCH-para} was synthesized as previously reported by our group \cite{liang2024dynamic}, and its precursors were synthesized as previously reported by Wimberger et al. \cite{wimberger2021large}.

\textsuperscript{1}H NMR (500 MHz, DMSO-d6) $\delta$ (ppm): 11.25 (s, 1H), 8.49 (d, \textit{J} = 16.0 Hz, 1H), 8.20 (d, \textit{J} = 8.9 Hz, 1H), 7.87 (d, \textit{J} = 8.9 Hz, 1H), 7.53 (d, \textit{J} = 2.5 Hz, 1H), 7.47 (d, \textit{J} = 16.1 Hz, 1H), 7.16 (dd, \textit{J} = 8.8, 2.5 Hz, 1H), 6.64 (dd, \textit{J} = 8.9, 2.5 Hz, 1H), 6.60 (d, \textit{J} = 2.4 Hz, 1H), 4.57 (t, \textit{J} = 7.8 Hz, 2H), 3.89 (s, 3H), 3.85 (s, 3H), 3.64 – 3.58 (m, 2H), 3.11 (s, 9H), 2.28 (t, \textit{J} = 4.4 Hz, 2H), 1.76 (s, 6H).

\textsuperscript{13}C NMR (126 MHz, DMSO) $\delta$ (ppm 179.63, 165.90, 161.45, 160.31, 147.71, 145.10, 134.07, 132.04, 115.65, 115.17, 114.65, 108.94, 108.23, 108.13, 100.67, 62.06, 56.20, 55.76, 52.55, 51.56, 43.11, 26.83, 21.73.

Spectral data matches those reported in the literature \cite{liang2024dynamic,wimberger2021large}.

\begin{figure*}[h!]
\centering
\includegraphics[width=\linewidth]{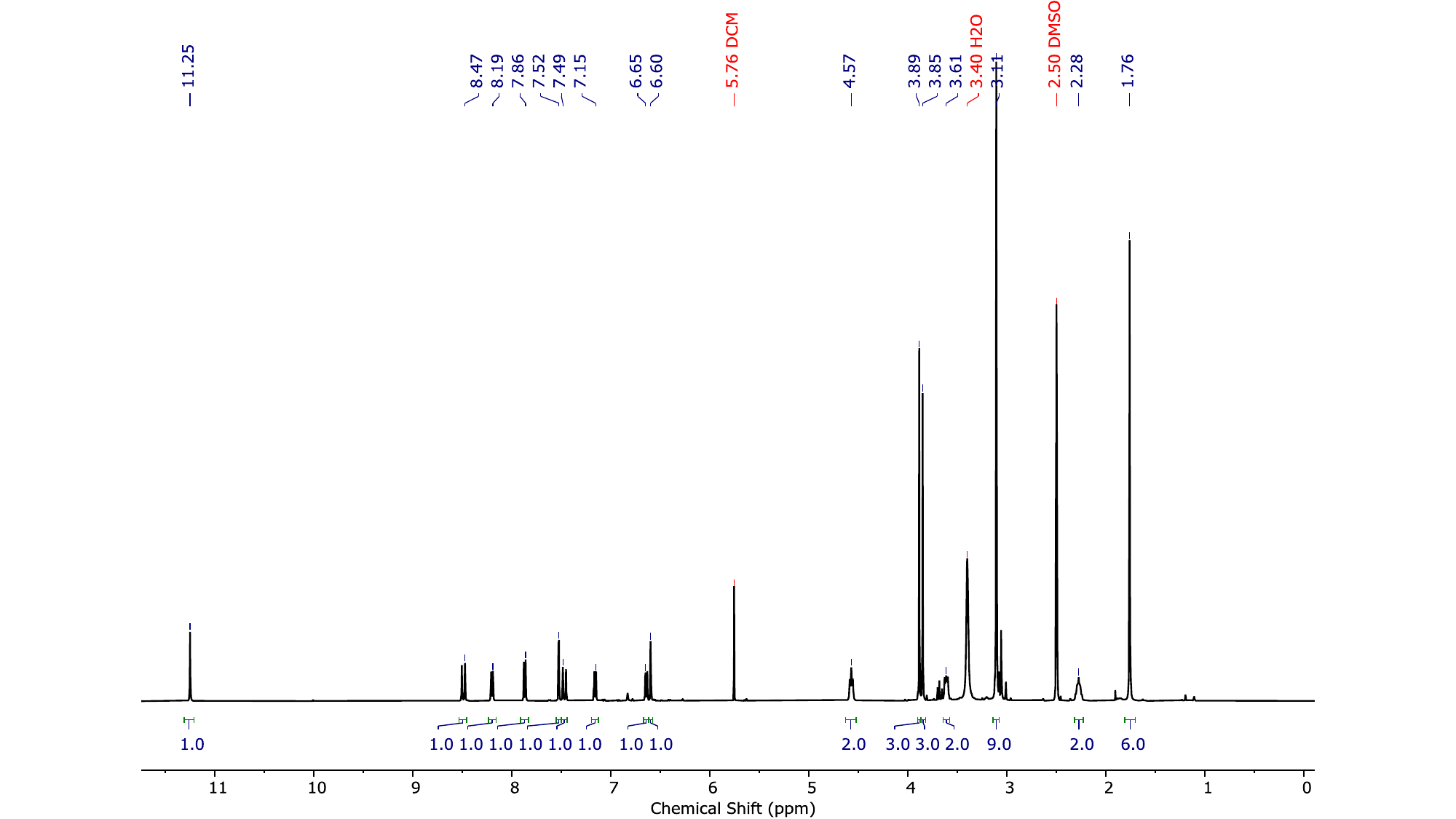}
\caption{\textsuperscript{1}H NMR spectrum of MCH-para in DMSO-\textit{d}6.}
\label{fig:S1}
\end{figure*}

\begin{figure*}[h!]
\centering
\includegraphics[width=\linewidth]{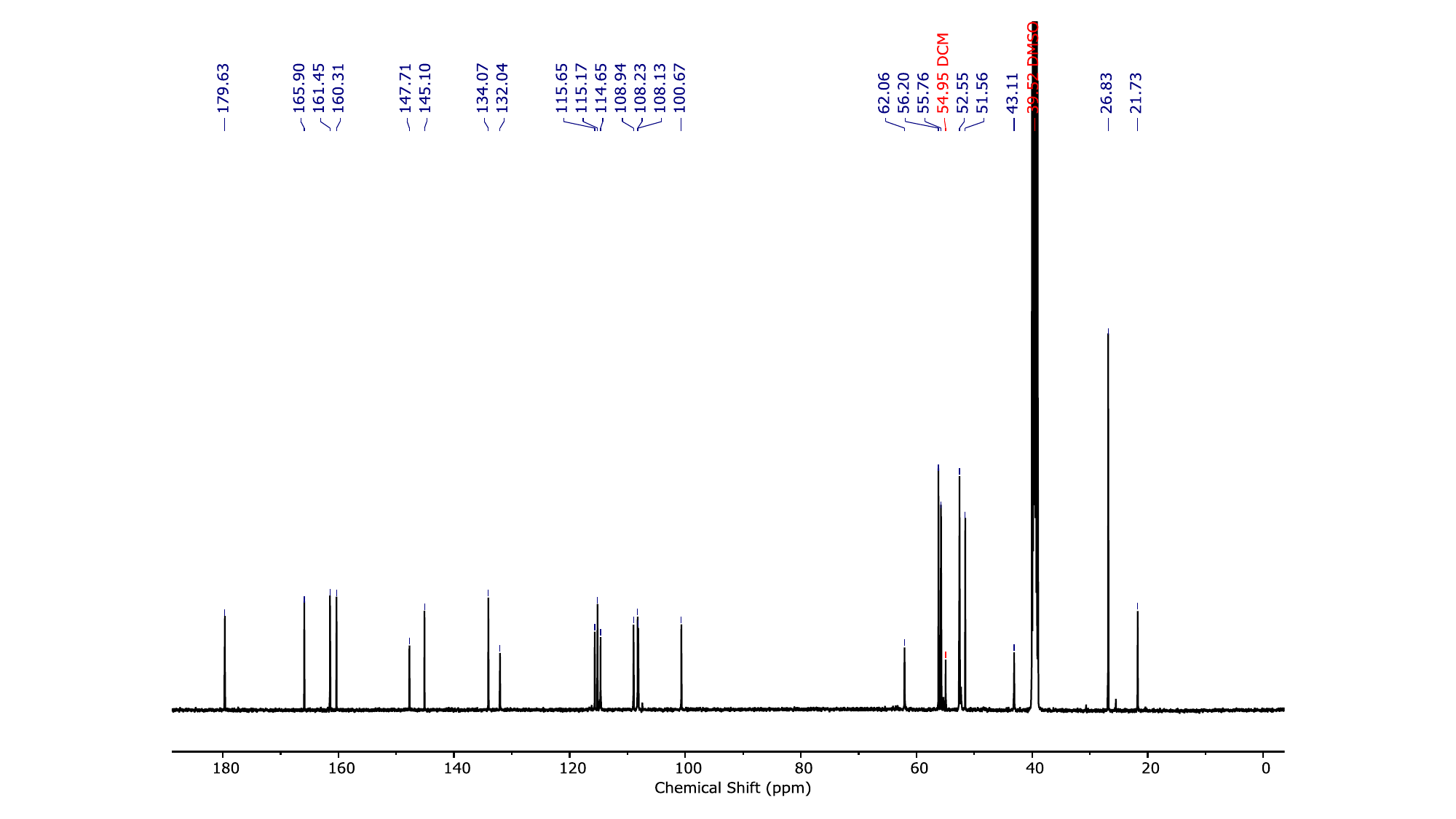}
\caption{\textsuperscript{13}C NMR spectrum of MCH-para in DMSO-\textit{d}6.}
\label{fig:S2}
\end{figure*}

\clearpage
\newpage

\subsection{DOSY experiment of MCH-para}
The diffusion coefficient of the MCH-para molecule in a dilute solution of CD\textsubscript{3}OD was quantified by DOSY experiments, which were performed on a Varian 600 MHz NMR spectrometer and processed with MestReNova software. These measurements were performed using the Varian DOSY pulse sequence Dbppste\_cc that encodes bipolar gradient pulses and stimulated echo with convection compensation \cite{wu1995improved,jerschow1997suppression}. Diffusion decays are collected with 18 diffusion experiments. \textit{I}\textsubscript{g} is the intensity, \textit{I}\textsubscript{0} is the initial intensity of the selected resonance, g is the applied gradient strength, the gradient pulse duration $\delta$ = 2.5 ms, and the total diffusion time $\Delta$ = 47.4 ms.

\begin{figure*}[h!]
\centering
\includegraphics[width=0.45\linewidth]{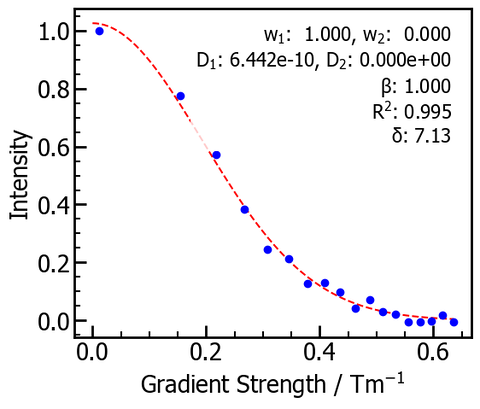}
\caption{DOSY plot showing the decay of signal intensity as a function of gradient strength (Tm$^{-1}$). The data points (blue dots) represent experimental NMR measurements, and the red dashed line indicates the fitted curve. The diffusion coefficients are calculated as $D_1 = 6.442 \times 10^{-10} \, \text{m}^2/\text{s}$. Fitting parameters include $w_1 = 1.000$, $w_2 = 0.000$, $\beta = 1.000$, $R^2 = 0.995$, and $\delta = 7.13$ ms.}
\label{fig: DOZY}
\end{figure*}

\begin{figure*}[h!]
\centering
\includegraphics[width=\linewidth]{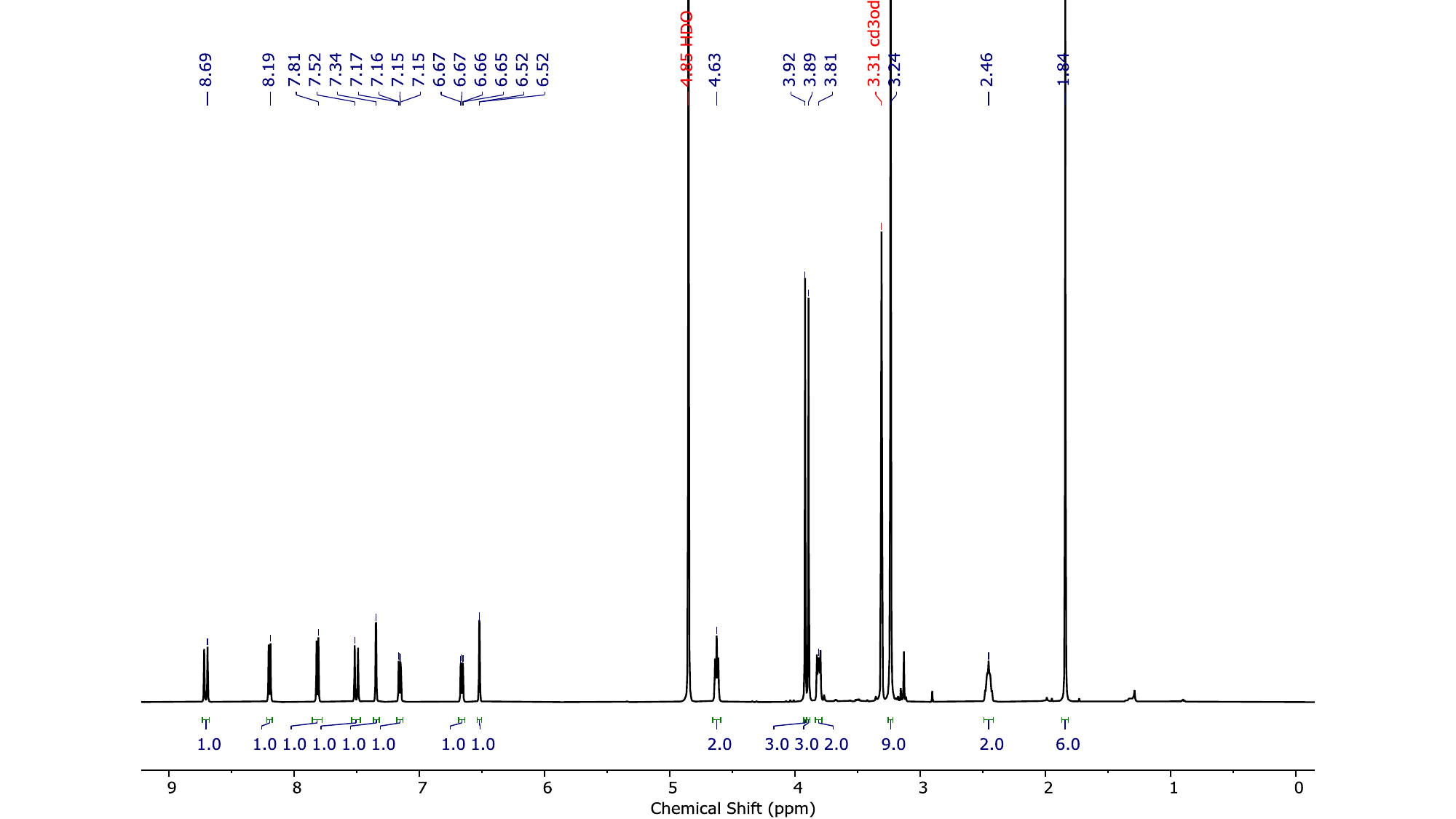}
\caption{\textsuperscript{1}H NMR spectrum of MCH-para in CD\textsubscript{3}OD}
\label{fig:S3}
\end{figure*}

\newpage
\subsection{Solution preparation for UV-Vis studies}\mbox{}
To comply with Beer-Lambert law, UV-Vis spectroscopy measurements were performed at a significantly lower concentration (0.025 mM) compared to the experimental setup used on the International Space Station (1 mM in MeOH solution). At reduced concentrations, merocyanine/spiropyran photoswitch may exhibit a complex equilibrium between multiple forms \cite{wimberger2021large}da. Specifically, the merocyanine photoswitch can exist in either a protonated form (MCH) or a deprotonated form (MC), depending on the solvent environment. At high concentrations of MCH-para (1 mM in MeOH), the protonated form is favored, displaying its characteristic red-orange color. However, when the solution was diluted below 0.05 mM in MeOH, a pink color, characteristic of the deprotonated MC form in protic solvents, was observed. To ensure consistent results at a 0.025 mM concentration for UV-Vis studies, we replicated the conditions of the MCH-para photosurfactant in the 1 mM MeOH solution (“pH” $\sim$3.5) by preparing the 0.025 mM solution in acidified MeOH. Under these conditions, the protonated MCH form, with its orange-red color, remains favored even at low concentrations. \textbf{Note:} For UV-Vis studies, it is essential that the total “pH” of the MCH-para solution in acidified MeOH remains between $\sim$2.5 and $\sim$3.5. If the solution is too acidic (“pH” $<$ 2), the photoswitching properties of MCH-para are reduced due to the equilibrium shifting significantly toward the MCH form.

\textbf{Acidification of Methanol.} The acidification of MeOH was performed as previously reported by our group \cite{chau2024photoresponsive}. 1.17 g (20 mmol) of NaCl was added to a 250 mL three-neck round-bottom flask equipped with an addition funnel and two rubber septa, followed by the slow addition of 1 mL of concentrated sulfuric acid. The HCl gas produced during the reaction was bubbled via a cannula into 100 mL of anhydrous MeOH in a round-bottom flask for 5 minutes. This flask was connected to an Erlenmeyer flask containing a dilute sodium bicarbonate solution, which acted as a trap for excess HCl gas. It is important to ensure the cannula is of adequate size to facilitate efficient HCl gas transfer. To achieve a consistent concentration of acid in methanol, the final “pH” of the MeOH was measured using a Thermo Scientific Orion Star™ A111 Benchtop pH Meter and subsequently diluted with non-acidified anhydrous MeOH until a “pH” of $\sim$2.5 was reached.

The stock solution of MCH-para for UV-Vis studies (0.1 mM) was prepared by dissolving 2.92 mg (0.005 mmol) of MCH-para in 50 mL of acidified MeOH. The solution was wrapped in aluminum foil and stored at 4 °C. Prior to conducting UV-Vis experiments, the stock solution was diluted to a 0.025 mM concentration, transferred to a quartz cuvette with a 1 cm path length, and placed in the spectrophotometer. The sample was allowed to equilibrate for 2 hours in the dark before measurement.

\clearpage
\newpage

\subsection{UV-Vis spectroscopy and UV-Vis kinetic measurements}
UV-Vis absorption spectra were recorded on an Agilent 8453 UV–Vis spectrometer from 200 to 1200 nm wavelengths. The photoinduced optical absorption kinetics was measured on a home-built pump-probe setup, as previously reported by our group \cite{hemmer2016tunable}. The blue pump beam was generated by a light-emitting diode (LED) source (Thorlabs, Inc. M470F3) coupled into a multimode optical fiber terminated with an output collimator. The LED intensity was controlled through a digital-to-analog converter (National Instruments USB-6009) using LabVIEW. The kinetic studies were conducted at 11 different light intensities. The probe beam was generated by an incandescent light bulb source (Ocean Optics LS1) coupled into a multimode fiber with an output collimator for light delivery. The probe light was modulated by a shutter (Uniblitz CS25), which could be controlled manually or through a digital output port (National Instruments USB-6009) using LabVIEW. Pump and probe beams overlapped using steering and focusing optics at a 90° angle inside a sample holder, allowing 10×10 mm\textsuperscript{2} rectangular spectrophotometer cells to be held within or cast film samples to be held to the front using metal spring clips. The solutions were continuously stirred during the measurements by a miniature stirring plate inserted into the sample holder (Starna Cells SCS 1.11). Both pump and probe beams were nearly collimated inside the cell with a diameter of about 2 mm. The pump beam was blocked after passing through the sample, and the probe beam was directed by a system of lenses into the detector (Ocean Optics Flame-S1-XR spectrometer), which acquired spectra of the probe light. The detector was connected to a PC via a USB port. The experiment was controlled by a National Instrument LabVIEW program, which collected the probe light spectra, determined sample optical absorption spectra, and controlled pump and probe light sources.

\subsection{Determining Reaction Rate Constants}
\begin{figure*}[h!]
\centering
\includegraphics[width=\linewidth]{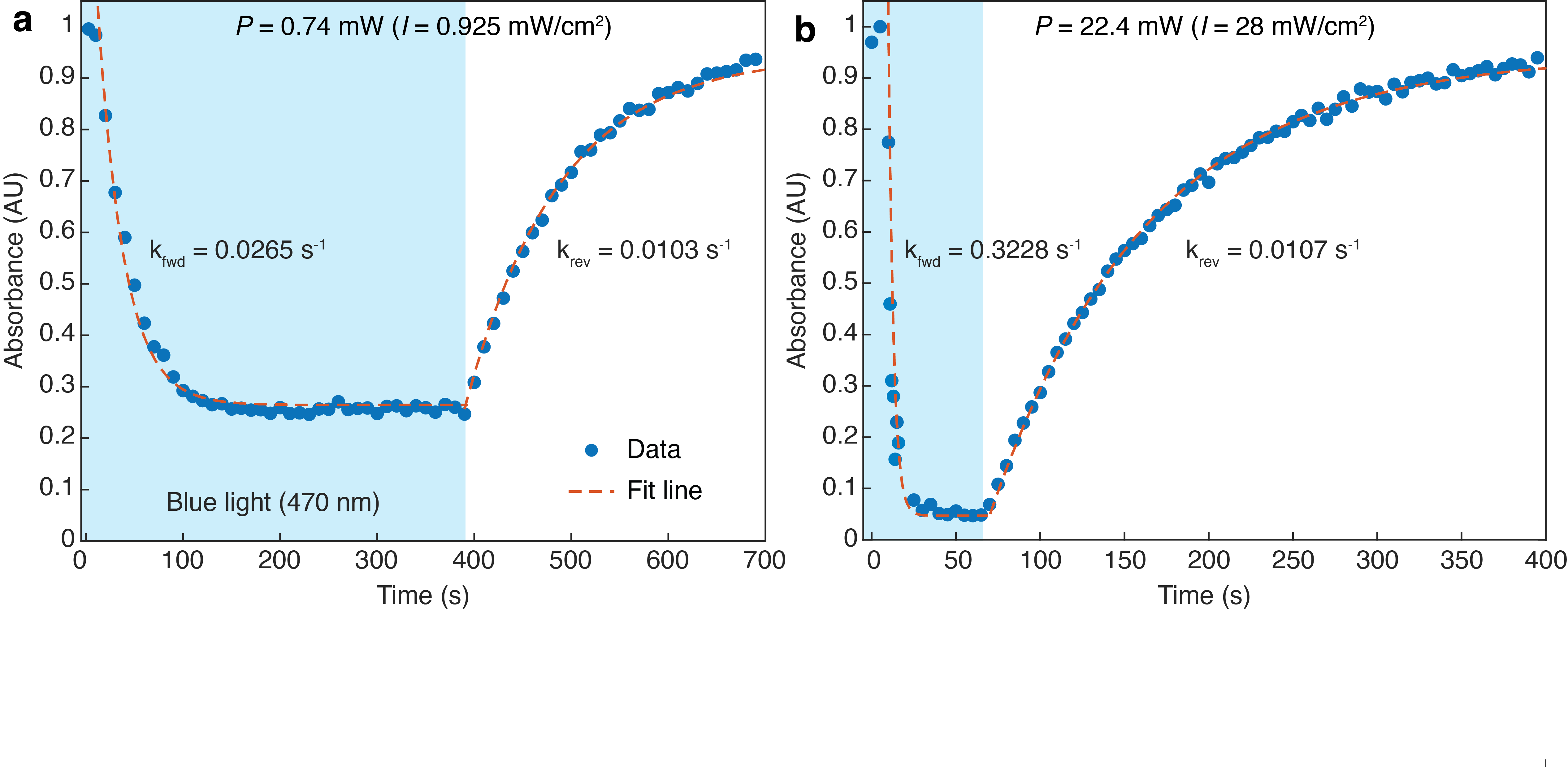}
\caption{\textbf{a.} Forward and reverse reaction fits for light power \( P = 0.74 \, \mathrm{mW} \) (intensity \( I = 0.925 \, \mathrm{mW/cm^2} \)) and \textbf{b.} \( P = 22.4 \, \mathrm{mW} \) (intensity \( I = 28 \, \mathrm{mW/cm^2} \)). Both data sets were modeled using first-order reaction kinetics. The shaded blue regions represent periods of 470 nm blue light exposure.
}
\label{fig: rate constant fit}
\end{figure*}

To model the kinetics of isomerization, we employed first-order reaction equations to fit the experimental data. The isomerization process, represented as $A \rightleftharpoons B$, was analyzed in both forward and reverse directions using light-induced switching. 

For the \textbf{reverse reaction}, the absorbance data was fitted with an exponential decay model:  
\[
A(t) = A_0 + (A_{\text{eq}} - A_0) \cdot \exp[-k_{\text{rev}} \cdot (t - t_{\text{off}})],
\]
where $A_0$ is the initial absorbance, $A_{\text{eq}}$ is the equilibrium absorbance, and $t_{\text{off}}$ is the time when the light is turned off. This model captures the decrease in absorbance as the system returns to equilibrium, with $k_{\text{rev}}$ representing the reverse rate constant.

For the \textbf{forward reaction}, we used a coupled model incorporating both forward and reverse rate dynamics:  
\[
A(t) = \frac{A_0}{k_{\text{fwd}} + k_{\text{rev}}} \left[ k_{\text{rev}} + k_{\text{fwd}} \cdot \exp[-(k_{\text{fwd}} + k_{\text{rev}}) \cdot (t - t_{\text{on}})] \right],
\]
where $k_{\text{fwd}}$ and $k_{\text{rev}}$ are the forward and reverse rate constants, respectively, and $t_{\text{on}}$ marks the time when the light was turned on. This equation reflects the competition between forward isomer conversion and reverse reversion processes under light exposure.

\begin{figure*}[h!]
\centering
\includegraphics[width=0.5\linewidth]{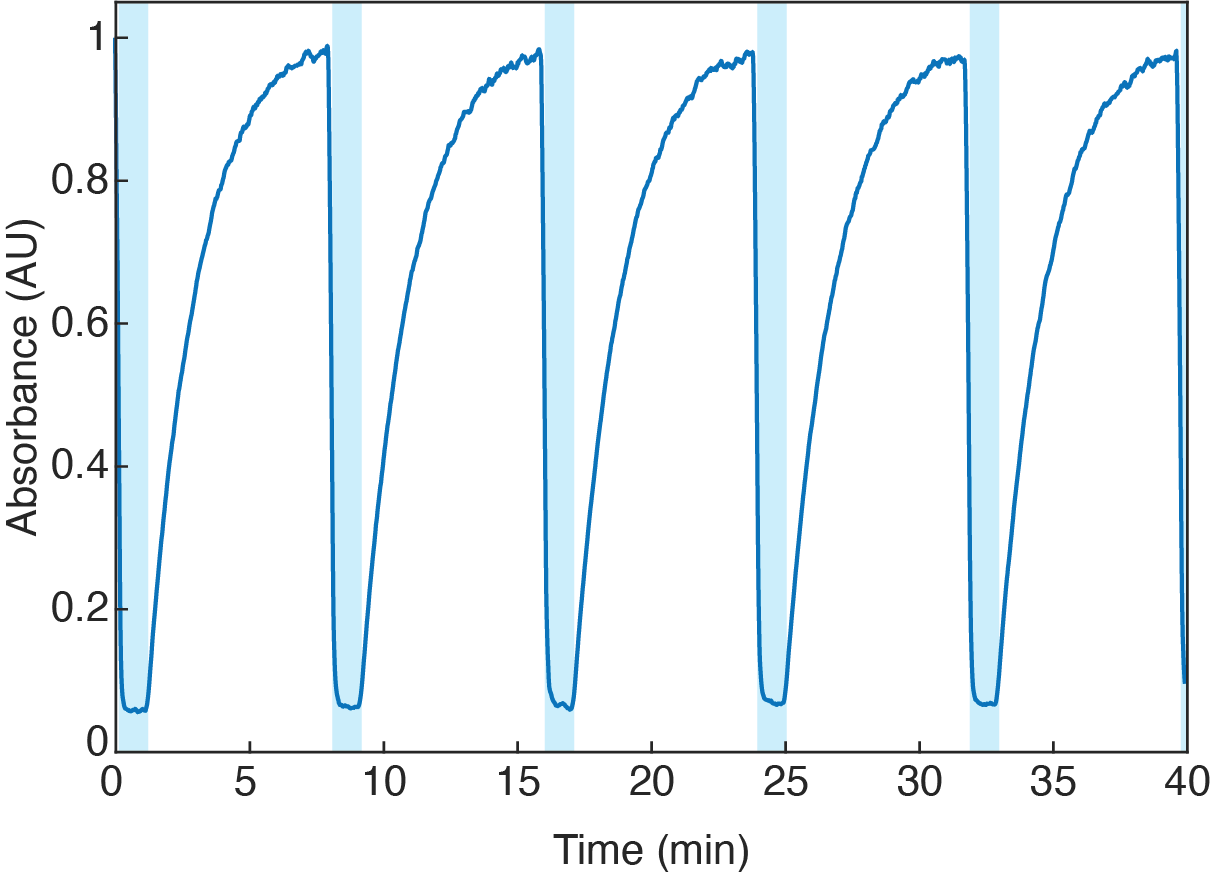}
\caption{Pump-probe kinetics measurements of MCH-para in acidified methanol (0.025 mM) irradiated with 470 nm light ($I = 2.8 \, \mathrm{mW/cm^2}$)}
\label{UV_Vis_5cycles}
\end{figure*}

\clearpage
\section{Surface Tension and Coverage Analysis}
\subsection{Impact of Background Lighting}
Surface tension in this study was measured using an optical tensiometer (Theta Flex, Biolin Scientific). This method relies on the pendant drop technique, where a droplet of the liquid sample is suspended from a syringe needle. A high-resolution camera captures the precise outline shape of the suspended drop under controlled lighting conditions. The acquired images are then analyzed using specialized software that applies the Young-Laplace equation to model the drop profile accurately.

\begin{figure*}[h!]
\centering
\includegraphics[width=\linewidth]{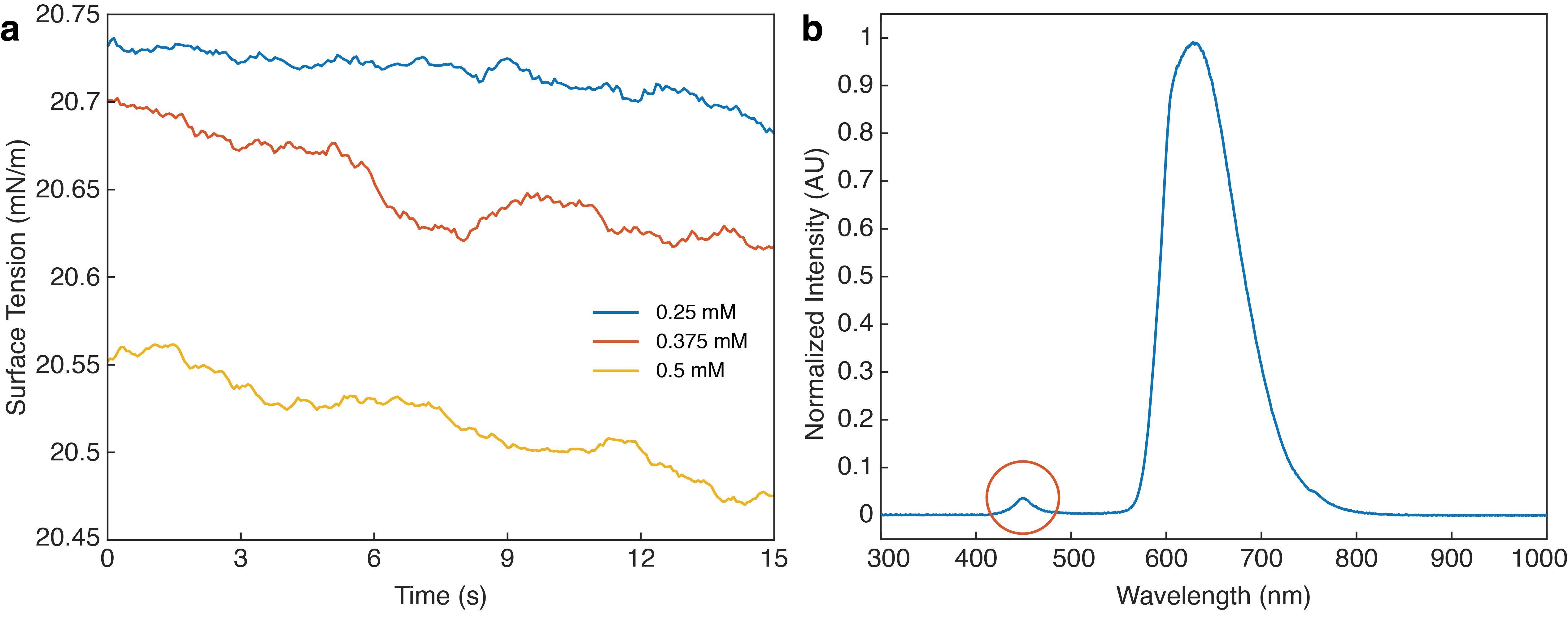}
\caption{\textbf{a.} Surface tension drop observed when the white backlight is turned on. \textbf{b.} Light intensity spectrum of the white light with a red optical filter, showing that the blue region retains intensity to induce a surface tension drop.}
\label{fig:backlight}
\end{figure*}

Due to the use of a camera, background lighting is necessary. However, the MCH-para molecules are sensitive to blue light, making it difficult to completely filter out blue light from the background illumination. To address this issue, we customized the background light by adding a transparent red filter to the white light, thereby reducing the transmission of blue light. Nevertheless, a small percentage of blue light still persists (Figure \ref{fig:backlight}a). As a result, the measured surface tension cannot fully revert to its original state under these lighting conditions. The surface tension measured using only the background light is presented in Figure \ref{fig:backlight}b.

\newpage
\subsection{Surface coverage analysis} 
Understanding the surface coverage of surfactants on the bubble surface is important for elucidating the behavior of the molecules and their impact on surface tension. This analysis involves modeling the adsorption and desorption kinetics of MCH and SP molecules (denoted as A and B in the equations for simplicity) and their interactions ($\tilde{A}$) at the interface. The following sections detail the mathematical framework and assumptions used to calculate surface coverage.

\textbf{Reversible reaction}
The adsorption and desorption of surfactant molecules are modeled as a reversible reaction between two species, A and B, as depicted in Equation \ref{eq:reaversible_reaction}:

\begin{equation}
\text{A} \stackrel{k_{\text{fwd}}}{\underset{k_{\text{rev}}}{\rightleftharpoons}} \text{B}
\label{eq:reaversible_reaction}
\end{equation}

where:\\
Species A represents the initial form of the surfactant, MCH, while Species B denotes the isomerized form, SP. $k_{\text{fwd}}$ and $k_{\text{rev}}$ are the forward and reverse rate constants for the conversion of A to B and B back to A, respectively.

The kinetics of this reversible reaction are governed by the rate equations, which describe the temporal changes in the concentrations of species A and B.
\begin{equation}
\frac{dC_A}{dt} = -k_{\text{fwd}} C_A + k_{\text{rev}} C_B = -\frac{dC_B}{dt}
\label{eq:reaction_rate}
\end{equation}

\textbf{Initial Conditions and Conservation of Mass} \\
At the beginning of the reaction, the system exclusively contains species A, with species B absent. This is mathematically represented by:
\begin{equation}
C_A = C_{A0}, \quad C_B = 0
\label{eq:initial_state}
\end{equation}

Conservation of mass dictates that the total concentration of surfactant remains constant throughout the reaction:

\begin{equation} 
C_A + C_B = C_{A0} 
\label{eq: mass conservation} 
\end{equation}

\textbf{Steady-State Concentrations}\\
At steady state, the concentrations of species A and B no longer change with time. Solving Equation \ref{eq:reaction_rate} under steady-state conditions yields:
\begin{equation}
C_A = \frac{k_{\text{rev}}}{k_{\text{rev}} + k_{\text{fwd}}} C_{A0}, \quad C_B = \frac{k_{\text{fwd}}}{k_{\text{rev}} + k_{\text{fwd}}} C_{A0}
\label{eq:steady_state}
\end{equation}

\textbf{Surface Tension and the Frumkin Isotherm}\\
Surface tension ($\gamma$) is influenced by the adsorption of surfactant molecules at the interface. The Frumkin isotherm provides a relationship between surface tension and the fractional surface coverage ($\theta$), accounting for interactions ($\tilde{A}$) between adsorbed molecules:

\begin{equation}
\gamma = \gamma_0 + {nRT} \cdot \Gamma_m [\ln (1-\theta) - \tilde{A} \frac{\theta^2}{2}]
\label{eq:Frumkin_isotherm}
\end{equation}

\noindent where:

\begin{description}
    \item[$\gamma_0$] is the surface tension of the pure solvent.
    
    \item[$n = 2$] for ionic surfactant.
    
    \item[$R$] is the universal gas constant.
    
    \item[$T$] is the absolute temperature.
    
    \item[$\Gamma_m$] is the maximum surface concentration of the surfactant.
    
    \item[$\theta = \dfrac{\Gamma_T}{\Gamma_m}$] is the fractional surface coverage, where $\Gamma_T$ is the total surface concentration of adsorbed surfactants.
    
    \item[$\tilde{A}$] is the interaction parameter accounting for lateral interactions between adsorbed molecules.
\end{description} 

\textbf{Adsorption and Desorption Dynamics}\\
The dynamic behavior of surfactant adsorption and desorption on the bubble surface is governed by the following set of differential equations:

\begin{equation}
\frac{d\Gamma_A}{dt} = k_{\text{rev}}\Gamma_B - k_{\text{fwd}}\Gamma_A + (\Gamma_m - \Gamma_T) \cdot \kappa_{\mathrm{ads}}^A C_A - \kappa_{\mathrm{des}}^A \Gamma_A \cdot e^{-\tilde{A} \frac{\Gamma_T}{\Gamma_m}}
\label{eq:surface_kinetics_A}
\end{equation}

\begin{equation}
\frac{d\Gamma_B}{dt} = k_{\text{fwd}}\Gamma_A - k_{\text{rev}}\Gamma_B + (\Gamma_m - \Gamma_T) \cdot \kappa_{\mathrm{ads}}^B C_B - \kappa_{\mathrm{des}}^B \Gamma_B \cdot e^{-\tilde{A} \frac{\Gamma_T}{\Gamma_m}}
\label{eq:surface_kinetics_B}
\end{equation}

whose sum yields:

\begin{equation}
\frac{d\Gamma_T}{dt} = (\Gamma_m - \Gamma_T) \cdot (\kappa_{\mathrm{ads}}^A C_A + \kappa_{\mathrm{ads}}^B C_B)- (\kappa_{\mathrm{des}}^A\Gamma_A + \kappa_{\mathrm{des}}^B \Gamma_B) \cdot e^{-\tilde{A} \frac{\Gamma_T}{\Gamma_m}}
\label{eq:surface_kinetics_Total}
\end{equation}

\noindent where:

\begin{description}
    \item[$\kappa_{\mathrm{ads}}$] is the adsorption rate constant.
    
    \item[$\kappa_{\mathrm{des}}$] is the desorption rate constant.
    
\end{description}

\noindent Adsorption Terms:

\begin{itemize}
    \item \((\Gamma_m - \Gamma_T) \cdot \kappa_{\text{ads}}^A C_A\) and \((\Gamma_m - \Gamma_T) \cdot \kappa_{\text{ads}}^B C_B\) represent the rate at which species A and B adsorb onto the surface, respectively.
    \item The factor \((\Gamma_m - \Gamma_T)\) ensures that adsorption is limited by the available surface sites.
\end{itemize}

\noindent Desorption Terms:

\begin{itemize}
    \item \(\kappa_{\text{des}}^A \Gamma_A \cdot e^{-\tilde{A} \frac{\Gamma_T}{\Gamma_m}}\) and \(\kappa_{\text{des}}^B \Gamma_B \cdot e^{-\tilde{A} \frac{\Gamma_T}{\Gamma_m}}\) account for the desorption of species A and B, respectively.
    \item The exponential term \(e^{-\tilde{A} \frac{\Gamma_T}{\Gamma_m}}\) incorporates the effect of lateral interactions between adsorbed molecules.
\end{itemize}

\textbf{Steady-State Surface Coverage}\\
Under steady-state conditions and assuming similar desorption rates for both species ($\kappa_{des}^A \approx \kappa_{des}^B$), Equation \ref{eq:surface_kinetics_Total} simplifies to:

\begin{equation}
0 = (\Gamma_m - \Gamma_T) \cdot (\kappa_F^A C_A + \kappa_F^B C_B)- (\Gamma_A + \Gamma_B) \cdot e^{-\tilde{A} \frac{\Gamma_T}{\Gamma_m}}
\label{eq:surface_kinetics_ss}
\end{equation}

Expressing this in terms of the fractional coverage $\theta$, the steady-state equation becomes:
\begin{equation}
0 = (1-\theta) \cdot (\kappa_F^A C_A + \kappa_F^B C_B)- \theta \cdot e^{-\tilde{A} \theta}
\label{eq:surface_kinetics_ss_theta}
\end{equation}

\textbf{Steady-State Coverage Under Different Light Conditions}\\
The surface coverage $\theta$ is influenced by illumination. The system can exist in two distinct steady states:

\paragraph{Dark State:} When the system is not illuminated, the coverage is governed solely by the adsorption of species A. The fractional coverage in the dark state is described by:
\begin{equation}
    \theta = \frac{\kappa_F^A C_{A0}}{\kappa_F^A C_A + e^{-\tilde{A} \theta}}
    \label{eq:surface_kinetics_ss_theta_dark}
\end{equation}

\paragraph{Illuminated State:} Under illumination, both forward and reverse reactions are active, altering the surface coverage:
\begin{equation}
    \theta = \frac{\kappa_F^A C_A + \kappa_F^B C_B}{\kappa_F^A C_A + \kappa_F^B C_B + e^{-\tilde{A} \theta}}
    \label{eq:surface_kinetics_ss_theta_light}
\end{equation}

\textbf{Calculation Procedure}\\
The surface coverage and surface tension were determined through a systematic computational approach implemented in MATLAB. The concentrations of the solution were defined and adjusted based on a conversion factor (derived from UV-Vis spectrum) to account for the interconversion between MCH and SP. Surface tension measurements were collected under both dark and light conditions (Figure 1c). These measurements were combined to facilitate a joint nonlinear regression analysis using the Frumkin isotherm model, to describe the dependence of surface tension on fractional coverage. 

The regression process involved defining an initial set of parameter guesses, including $\kappa_F^A$ and $\kappa_F^B$, $\tilde{A}$, $\Gamma_m$, and $\gamma_0$. A joint model function was constructed to simultaneously fit the dark and light condition data, leveraging separate functions for each state. The \texttt{fitnlm} function in MATLAB was employed to perform the nonlinear regression, optimizing the model parameters to best fit the experimental data.

\begin{figure*}[h!]
\centering
\includegraphics[width=0.5\linewidth]{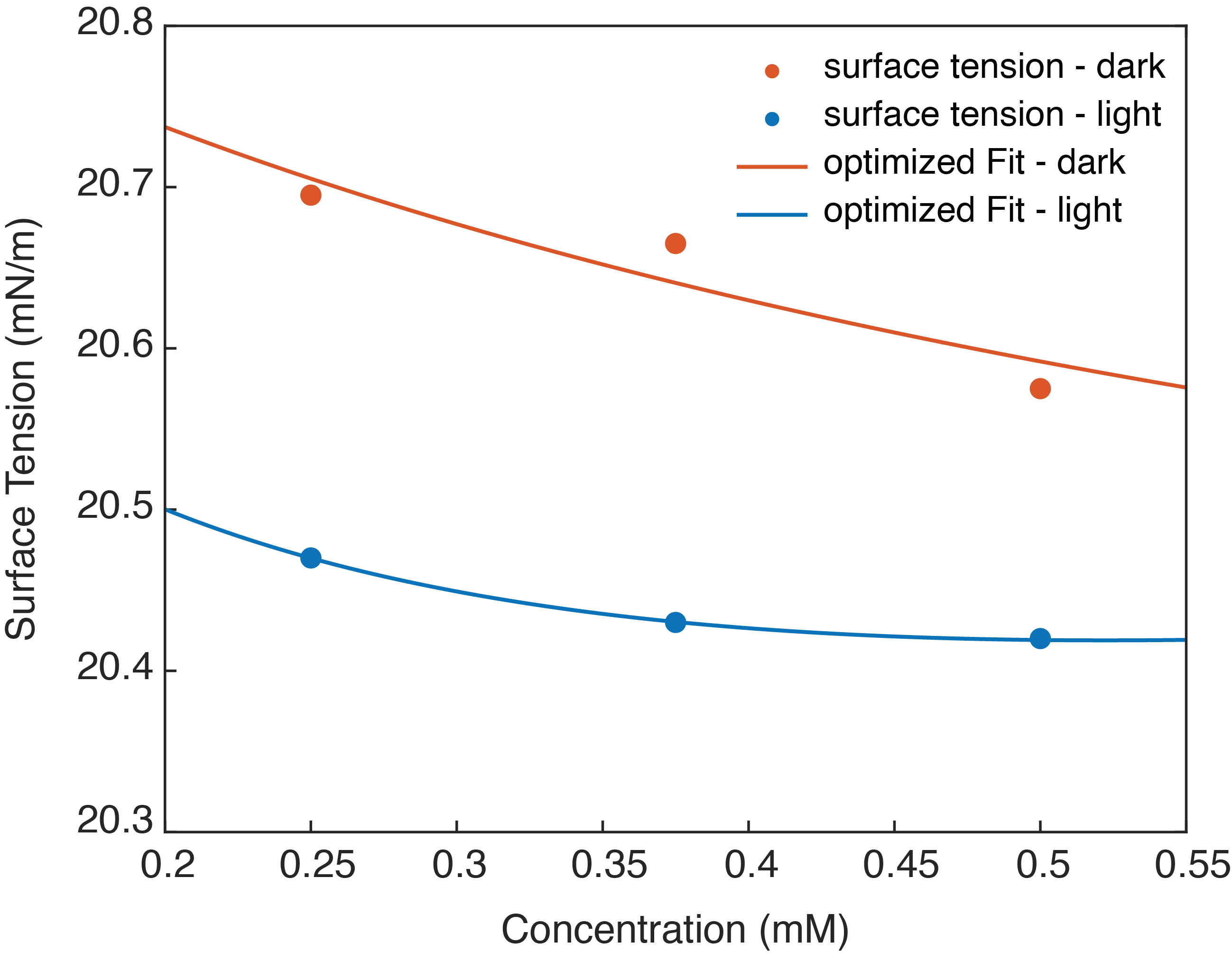}
\caption{Surface tension measurements at different concentrations, with the fitted curve computed using the Frumkin isotherm to model the corresponding surface tension.}
\label{matlab:optimized parameters}
\end{figure*}

Subsequently, the optimized parameters were extracted to calculate the fractional surface coverage (\(\theta\)) for both dark and light conditions. This calculation utilized iterative numerical methods to solve the nonlinear equations derived from the Frumkin isotherm. The optimized \(\theta\) values were then substituted back into the Frumkin isotherm to compute the corresponding surface tension (\(\gamma\)) (Figure \ref{matlab:optimized parameters}).

\textbf{Surface Tension Measurements and Numerical Solutions for Surface Coverage}\\
Optimized Parameters:

\begin{description}
    \item[$\kappa_F^A$]  = 0.070029 m\textsuperscript{3}/mol
    \item[$\kappa_F^B$] = 0.322124 m\textsuperscript{3}/mol
    \item[$\tilde{A}$] = $-19.321002$
    \item[$\Gamma_m$] = $3.74 \times 10^{-6}$ mol/m\textsuperscript{2}
    \item[$\gamma_0$] = 20.926 mN/m
\end{description}

\vspace{1cm} 
\begin{table}[h]
\captionsetup{justification=raggedright, singlelinecheck=false} 
\begin{tabular}{|c|c|c|c|c|}
\hline
\makecell{\textbf{Concentration (mM)}} & 
\makecell{\textbf{$\gamma_{dark}$ (mN/m)}} & 
\makecell{\textbf{$\theta_{dark}$}} & 
\makecell{\textbf{$\gamma_{light}$ (mN/m)}} &
\makecell{\textbf{$\theta_{light}$}} \\
\hline
0.25 & 20.70 & 0.0136 & 20.47 & 0.0374 \\
\hline
0.375 & 20.67 & 0.0186 & 20.43 & 0.0465 \\
\hline
0.5 & 20.58 & 0.0228 & 20.42 & 0.0536 \\
\hline
\end{tabular}
\caption{Surface tension and surface coverage at different concentrations.}
\label{T1}
\end{table}

The depletion length ($L_d$) is a characteristic length scale that quantifies the extent to which surfactant molecules are depleted from the bulk solution near the liquid-air boundary. It is defined as:

\begin{equation}
L_d = \Gamma_m \times \kappa
\label{eq:depletion_length}
\end{equation}

By using the model fitted to the measurements in various trials, even though there exist variations in the $\Gamma_m$ and $\kappa$ for molecules MCH and SP separately, the depletion length, calculated by multiplying both factors, consistently remains comparable in magnitude (262 nm for MCH and 1204 nm for SP), approximately . This consistency implies that the region adjacent to the surface, where the concentration of the surfactants is significantly reduced, maintains a uniform spatial extent. 

\newpage
\textbf{Time-Dependent Concentration Dynamics}\\

\begin{equation}
\frac{dC_A}{dt} = - k_{\text{fwd}} C_A + k_{\text{rev}} C_B
\end{equation}

\begin{equation}
C_A + C_B = C_{A0}
\end{equation}

\begin{equation}
\frac{dC_A}{dt} = - k_{\text{fwd}} C_A + k_{\text{rev}} (C_{A0} - C_A)
\end{equation}

\begin{equation}
\frac{dC_A}{dt} = -(k_{\text{rev}} + k_{\text{fwd}}) C_A + k_{\text{rev}} C_{A0}
\end{equation}

Multiplying by \( e^{(k_{\text{fwd}} + k_{\text{rev}})t} \):

\begin{equation}
e^{(k_{\text{fwd}} + k_{\text{rev}})t} \frac{dC_A}{dt} + e^{(k_{\text{fwd}} + k_{\text{rev}})t} (k_{\text{rev}} + k_{\text{fwd}})C_A = e^{(k_{\text{fwd}} + k_{\text{rev}})t} k_{\text{rev}} C_{A0}
\end{equation}

Using the product rule:

\begin{equation}
\frac{d}{dt} \left[ C_A \cdot e^{(k_{\text{fwd}} + k_{\text{rev}})t} \right] = e^{(k_{\text{fwd}} + k_{\text{rev}})t} \cdot k_{\text{rev}} \cdot C_{A0}
\end{equation}

Integrating both sides:

\begin{equation}
\int d \left[ C_A \cdot e^{(k_{\text{fwd}} + k_{\text{rev}})t} \right] = \int k_{\text{rev}} \cdot C_{A0} \cdot e^{(k_{\text{fwd}} + k_{\text{rev}})t} dt
\end{equation}

Solving the integral:

\begin{equation}
C_A \cdot e^{(k_{\text{fwd}} + k_{\text{rev}})t} = \frac{C_{A0} \cdot k_{\text{rev}}}{k_{\text{fwd}} + k_{\text{rev}}} \cdot e^{(k_{\text{fwd}} + k_{\text{rev}})t} + C
\end{equation}

Applying the initial condition \( C_A(0) = C_{A0} \):

\begin{equation}
C_{A0} = \frac{C_{A0} \cdot k_{\text{rev}}}{k_{\text{fwd}} + k_{\text{rev}}} + C
\end{equation}

Solving for \( C \):

\begin{equation}
C = \left(1 - \frac{k_{\text{rev}}}{k_{\text{fwd}} + k_{\text{rev}}} \right) \cdot C_{A0}
\end{equation}

Thus, the final expression for \( C_A(t) \) is:

\begin{equation}
C_A(t) = \frac{k_{\text{rev}}}{k_{\text{fwd}} + k_{\text{rev}}} \cdot C_{A0} + \frac{k_{\text{fwd}}}{k_{\text{fwd}} + k_{\text{rev}}} \cdot C_{A0} \cdot e^{-(k_{\text{fwd}} + k_{\text{rev}})t}
\end{equation}

Given the rate constants:

\[
k_{\text{fwd}} = 0.4875 \, \text{s}^{-1}, \quad k_{\text{rev}} = 0.0106 \, \text{s}^{-1}
\]

The general equation for \( C_A(t) \) is:

\begin{equation}
C_A(t) = \frac{k_{\text{rev}}}{k_{\text{fwd}} + k_{\text{rev}}} \cdot C_{A0} + \frac{k_{\text{fwd}}}{k_{\text{fwd}} + k_{\text{rev}}} \cdot C_{A0} \cdot e^{-(k_{\text{fwd}} + k_{\text{rev}})t}
\end{equation}

Substituting the given values:

\begin{equation}
C_A(t) = \frac{0.0106}{0.4875 + 0.0106} \cdot C_{A0} + \frac{0.4875}{0.4875 + 0.0106} \cdot C_{A0} \cdot e^{- (0.4875 + 0.0106) t}
\end{equation}

\begin{equation}
C_A(t) = \frac{0.0106}{0.4981} \cdot C_{A0} + \frac{0.4875}{0.4981} \cdot C_{A0} \cdot e^{-0.4981 t}
\end{equation}

Approximating the fractions:

\begin{equation}
C_A(t) \approx 0.0213 \cdot C_{A0} + 0.9787 \cdot C_{A0} \cdot e^{-0.4981 t}
\end{equation}

\noindent
Substituting given values:

\begin{equation}
C_A(2 s) \approx 0.3827, \quad C_B(2 s) \approx 0.6173
\end{equation}

\begin{equation}
C_A(10 s) \approx 0.0280, \quad C_B(10 s) \approx 0.9720
\end{equation}

\clearpage
\newpage
\section{Estimation of Capillary Length, Buoyancy and Marangoni Forces}
\begin{table}[h]
\centering
\begin{tabular}{|c|l|c|c|}
\hline
\textbf{Property} & \textbf{Description}                              & \textbf{Value (30 °C)}       & \textbf{Unit} \\ \hline
$g$             & Acceleration due to gravity                       & 9.81                 & $\text{m/s}^2$ \\ \hline
$\beta$         & Thermal expansion coefficient                     & 0.00154                    & $\text{K}^{-1}$ \\ \hline
$\Delta T$      & Temperature difference across the fluid           & 0.3                    & $\text{K}$ \\ \hline
$\rho$          & Density of the fluid                              & 785                    & $\text{kg/m}^3$ \\ \hline
$c_p$           & Specific heat capacity at constant pressure       & 2.53                    & J/(kg $\cdot$ K) \\ \hline
$L_c$           & Characteristic length scale                       & $1.65 \times 10^{-3}$ & $\text{m}$ \\ \hline
$\mu$           & Dynamic viscosity of the fluid                  & 0.0005                    & $\text{Pa} \cdot \text{s}$ \\ \hline
$\nu$           & Kinematic viscosity of the fluid                  & $6.37 \times 10^{-7}$  & $\text{m}^2/\text{s}$ \\ \hline
$k$             & Thermal conductivity                              & 0.202                    & W/(m $\cdot$ K) \\ \hline
\end{tabular}
\caption{Parameters, their values, and units }
\end{table}

\textbf{Capillary Length:} \\
\begin{equation}
L_c = \sqrt{\frac{\gamma}{\rho g}} = \sqrt{\frac{0.0212 \, \text{N/m}}{791 \, \text{kg/m}^3 \times 9.81 \, \text{m/s}^2}} 
= \sqrt{\frac{0.0212}{7754.71}} \approx \sqrt{2.734 \times 10^{-6}} \approx 0.001653 \, \text{m} \approx 1.65 \, \text{mm}
\label{eq:capillary_length}
\end{equation}

\textbf{Buoyancy Force:} \\
The volume \(V\) of a bubble with radius \(r = 1.65 \, \text{mm}\) is given by:

\begin{equation}
V = \frac{4}{3} \pi r^3 = \frac{4}{3} \pi (0.00165 \, \text{m})^3 \approx 1.88 \times 10^{-8} \, \text{m}^3
\label{eq:bubble_volume}
\end{equation}

The buoyancy force \(F_b\) acting on the bubble is given by:

\begin{equation}
F_b = \rho V g = (791 \, \text{kg/m}^3) \times \left(1.88 \times 10^{-8} \, \text{m}^3\right) \times 9.81 \, \text{m/s}^2 \approx 1.45 \times 10^{-4} \, \text{N}
\label{eq:buoyancy_force}
\end{equation}

\textbf{Marangoni Force:} \\
The mass of the displaced liquid can be calculated as:

\begin{equation}
m = \rho V_{\text{displaced}} = \rho \times \frac{4}{3} \pi r^3 = 791 \, \text{kg/m}^3 \times \frac{4}{3} \pi (0.00165 \, \text{m})^3 \approx 1.49 \times 10^{-5} \, \text{kg}
\label{eq:mass_displaced}
\end{equation}

Using the calculated mass and the given acceleration \( a = 7.6 \, \text{mm/s}^2 = 0.0076 \, \text{m/s}^2 \), the Marangoni force is estimated using:

\begin{equation}
F_{\text{M}} = m a = (1.49 \times 10^{-5} \, \text{kg}) \times (0.0076 \, \text{m/s}^2) \approx 1.13 \times 10^{-7} \, \text{N}
\label{eq:marangoni_force}
\end{equation}

\textbf{Buoyancy force due to Thermal expansion:} \\
The thermal expansion coefficients of methanol at \(20^\circ\mathrm{C}\) and \(40^\circ\mathrm{C}\) are \(\alpha_{20^\circ\mathrm{C}} = 0.00149\, \mathrm{^\circ C^{-1}}\) and \(\alpha_{40^\circ\mathrm{C}} = 0.00159\, \mathrm{^\circ C^{-1}}\), respectively. By linear extrapolation, the coefficient at \(30^\circ\mathrm{C}\) is:

\[
\alpha_{30^\circ\mathrm{C}} = 0.00154\, \mathrm{^\circ C^{-1}}.
\]

The total volume of the solution with a sealed air bubble of radius \(1.65\,\mathrm{mm}\) is calculated by subtracting the bubble's volume from the container volume:

\[
V_0 = (40 \times 10 \times 10\, \mathrm{mm}^3) - \frac{4}{3}\pi (1.65\,\mathrm{mm})^3 \approx 3981.18 \, \mathrm{mm}^3.
\]

Given the density of methanol is \(785 \, \mathrm{kg/m}^3\), the mass of the solution is:

\[
\text{Mass} = V_0 \times \rho = 3.98118 \times 10^{-6} \, \mathrm{m}^3 \times 785 \, \mathrm{kg/m}^3 = 3.123 \, \mathrm{g}.
\]

An increase of \(0.3^\circ\mathrm{C}\) results in a volume change due to thermal expansion:

\[
\Delta V = V_0 \cdot \beta \cdot \Delta T = 3981.18 \, \mathrm{mm}^3 \times 0.00154 \, \mathrm{^\circ C^{-1}} \times 0.3 \, \mathrm{^\circ C} = 1.84 \, \mathrm{mm}^3,
\]

yielding a new volume:

\[
V_{\text{new}} = 3981.18 \, \mathrm{mm}^3 + 1.84 \, \mathrm{mm}^3 = 3983.02 \, \mathrm{mm}^3.
\]

The new density is:

\[
\rho_{\text{new}} = \frac{\text{mass}}{V_{\text{new}}} = \frac{3.123 \, \mathrm{g}}{3.98302 \, \mathrm{cm}^3} \approx 0.784 \, \mathrm{g/cm}^3 = 784 \, \mathrm{kg/m}^3.
\]

The buoyancy force \( F_b \) rise is given by:

\[
\Delta F_b = \Delta \rho \cdot V \cdot g,
\]

where \(\Delta \rho = 785 \, \mathrm{kg/m}^3 - 784 \, \mathrm{kg/m}^3 = 1 \, \mathrm{kg/m}^3\), \( V = \frac{4}{3} \pi (0.00165 \, \mathrm{m})^3 = 1.88 \times 10^{-8} \, \mathrm{m}^3\), and \(g \approx 9.81 \, \mathrm{m/s}^2\). Substituting these values:

\[
F_b = 1 \, \mathrm{kg/m}^3 \times 1.88 \times 10^{-8} \, \mathrm{m}^3 \times 9.81 \, \mathrm{m/s}^2 = 1.84 \times 10^{-7} \, \mathrm{N}.
\]

Therefore, considering thermal expansion, the buoyancy force acting on the bubble is approximately \(1.446 \times 10^{-4}\) N.

\clearpage
\section{Thermal effect}
\captionsetup{justification=raggedright, singlelinecheck=false} 
\begin{figure*}[h!]
\centering
\includegraphics[width=\linewidth]{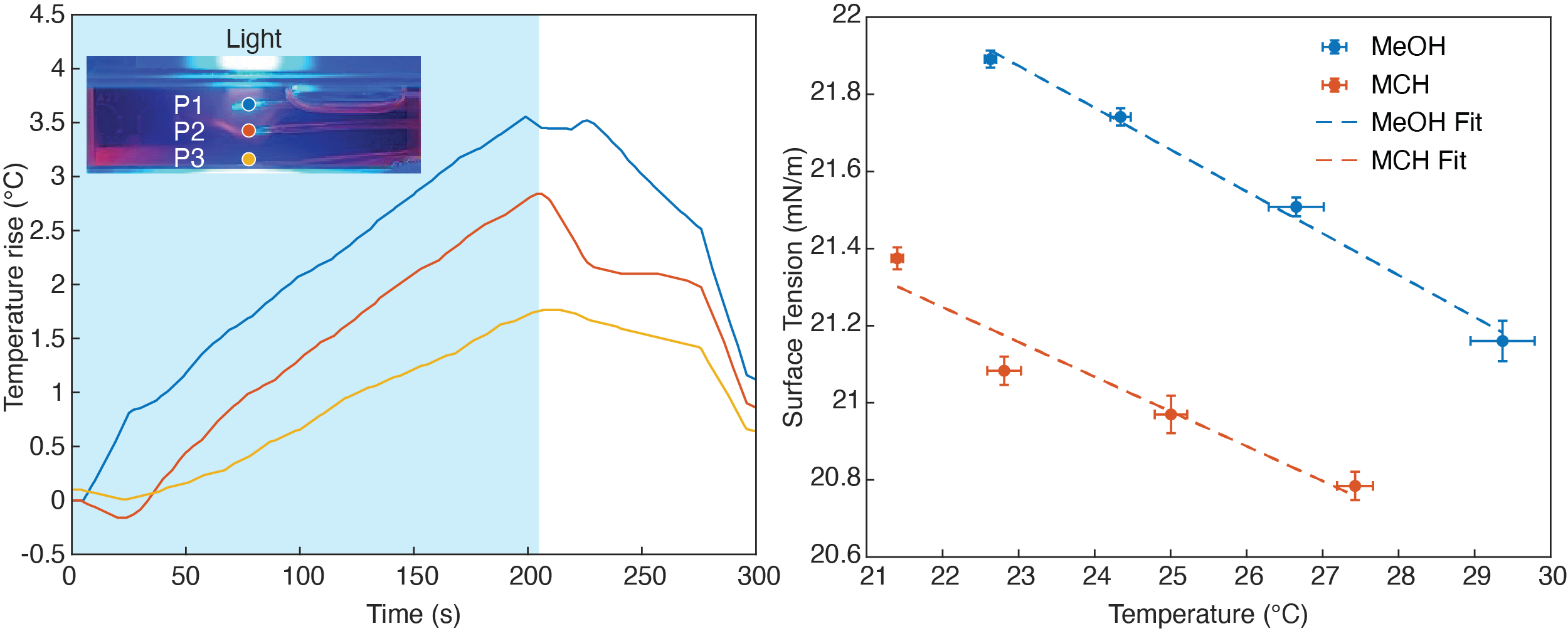}
\caption{\justifying
\textbf{a.} Temperature measurements at three positions over time under illumination from three 470 nm LEDs arranged in a row at 37.5 mW/cm\textsuperscript{2}.
\textbf{b.} Surface tension response of pure methanol and 1 mM MCH-para in methanol measured at temperatures ranging from 21 °C to 30 °C.}
\label{Fig. thermal effect}
\end{figure*}

The thermal-Marangoni effect can occur during bubble migration when the liquid absorbs light and heats up, resulting in an uneven temperature distribution along the bubble's surface. To estimate the temperature rise in space experiments, we first measured the temperature using thermocouples placed at three positions (P1, P2, and P3) from top to bottom of the setup, as shown in the inset images of Figure \ref{Fig. thermal effect}a. In this experiment, a 1 mM MCH solution in methanol was sealed in a four-sided polished quartz cuvette at room temperature, replicating the conditions used in space experiments. Three LEDs (470 nm, 37.5 mW/cm\textsuperscript{2}) were arranged in a row above the cuvette for illumination, and the temperature was continuously recorded after the light was turned on. As shown in the figure, the highest temperature rise observed was 3.5°C after more than 200 seconds of heating. In the space experiments, the longest heating duration is 10 seconds, generating a temperature rise of less than 0.18°C.

The change in surface tension due to heating was evaluated by measuring the surface tension of pure methanol and a 1 mM MCH methanol solution at temperatures ranging from 21 °C to 30 °C, as shown in Figure \ref{Fig. thermal effect}b. For the 1 mM MCH methanol solution (in orange), the change in surface tension during heating is described by the following equation:

\begin{equation}
\gamma (mN/m) = -0.09 \cdot T + 23.23
\label{eq:T_rise}
\end{equation}

With a 0.18°C change in temperature, the surface tension decreases by approximately 0.016 mN/m, which is significantly less than the surface tension change caused by the photo-Marangoni effect, which ranges from 0.14 to 0.27 mN/m (Figure 1c). Therefore, even when the thermal effect coexists with the photo effect, its contribution is minimal.

\clearpage
\section{Experimental conditions}

\begin{table}[h]
\centering
\begin{tabular}{|c|c|c|}
\hline
\textbf{Trial} & \textbf{Bubble Radius (mm)} & \textbf{Light Duration (s)} \\ \hline
1 & 3.053 & 10 \\ \hline
2 & 2.425 & 10 \\ \hline
3 & 2.390 & 10 \\ \hline
4 & 0.981 & 10 \\ \hline
5 & 3.050 & 8  \\ \hline
6 & 2.448 & 8  \\ \hline
7 & 2.399 & 8  \\ \hline
8 & 3.060 & 5  \\ \hline
9 & 2.484 & 5  \\ \hline
10 & 2.440 & 5 \\ \hline
11 & 3.030 & 2 \\ \hline
12 & 2.525 & 2 \\ \hline
13 & 2.466 & 2 \\ \hline

\end{tabular}
\caption{Experimental conditions of bubble size and light duration. Each trial presents the bubble radius and corresponding light duration, with 15-20 segments conducted under identical conditions for each trial.}
\label{tab: Experimental conditions}
\end{table}

\begin{table}[h]
\centering
\begin{tabular}{|c|c|c|c|}
\hline
  & Relative humidity \%H & Temperature °C (RTD) \\ \hline
Average   & 16.62              & 29.31               \\ \hline
Std Dev   & 3.15              & 0.843         \\ \hline
\end{tabular}
\caption{Humidity and temperature readings of CubeLab. The table shows average values and standard deviations of humidity from a PCB-mounted sensor and temperature from an ambient RTD sensor. }
\label{tab:Humidity and Temperature in CubeLab}
\end{table}

\clearpage
\section{COMSOL Simulation}
\subsection{Key physical concepts and governing equations}
\textbf{Chemical Reaction}: There is a reversible reaction between two surfactant forms, MCH and SP, driven by light (photoswitching) and from SP to MCH (thermal relaxation). The forward reaction from MCH to SP and the reverse reaction from SP to MCH are controlled by reaction rate constants $k_{\text{fwd}}$ and $k_{\text{rev}}$, respectively.

  \[
  \text{Forward Reaction Rate (from MCH to SP):} \quad \text{Rate} = k_{\text{fwd}} \, C_{\text{MCH}}
  \]

  \[
  \text{Reverse Reaction Rate (from SP to MCH):} \quad \text{Rate} = k_{\text{rev}} \; C_{\text{SP}}
  \]

\textbf{Bulk Advection and Diffusion}: The bulk surfactant concentration $C_i(x, y)$ undergoes transport in the bulk, influenced by fluid velocity $u$ and diffusion coefficient $D$. Bulk advection is governed by the term $-\nabla \cdot (\vec{u} C_i)$ for advection, while diffusion is described by $D \nabla^2 C_i$.

  \[
  \text{Bulk Advection and Diffusion:} -\nabla \cdot (\vec{u} C_i) + D \nabla^2 C_i
  \]

\textbf{Surface Transport, advection and diffusion}: Surfactant molecules can also advect along the bubble surface, described by $-\nabla_s \cdot (\vec{u} \Gamma_i)$, where $\Gamma_i(x, y)$ represents the surface concentration of surfactant, and $\nabla_s$ is the surface gradient operator. Surface diffusion is $D_s \nabla^2 \Gamma_i$, where $D_s$ is the surface diffusivity.

  \[
  \text{Surface Transport:} \quad -\nabla_s \cdot (\vec{u}_s \Gamma_i) + D_s \nabla^2 \Gamma_i
  \]

\textbf{Adsorption and Desorption}: Surfactant molecules can adsorb onto and desorb from the bubble surface. The rate of adsorption depends on the bulk concentration and the available surface area, while the rate of desorption depends on the current surface concentration.

  \[
  \text{Rate of Adsorption:} \quad (\Gamma_m - \Gamma_T) \;k_{\text{ads}}^i \; C_i
  \]

  \[
  \text{Rate of Desorption:} \quad k_{\text{des}}^i \; \Gamma_i \; e^{-\frac{A_i \Gamma_T}{\Gamma_m}}
  \]

\textbf{Surface Tension}: The surface tension $\gamma$ is affected by the surfactant concentration on the surface and is a key factor driving Marangoni flows along the bubble interface.

  \[
  \text{Surface Tension Equation:} \quad \gamma = \gamma_0 + nRT \; \ln \left(1 - \frac{\Gamma}{\Gamma_m}\right)
  \]

\textbf{Overall governing equations}: Once we include also the Navier-Stokes equations of fluid motion, the overall system of equations that is solved can be written as follows; in the bulk:
\begin{align}
   \rho \left(\frac{\partial \vec{u} }{\partial t} + \vec{u}\cdot\nabla \vec{u}\right) = -\nabla p + \mu \nabla^2 \vec{u}  \\
   \frac{\partial C_\text{SP} }{\partial t} + \vec{u}\cdot\nabla C_\text{SP}= D \nabla^2 C_\text{SP} +k_\text{fwd} C_\text{MCH} -k_\text{rev} C_\text{SP}  \\
   \frac{\partial C_\text{MCH} }{\partial t} + \vec{u}\cdot\nabla C_\text{MCH}= D \nabla^2 C_\text{MCH} -k_\text{fwd} C_\text{MCH} +k_\text{rev} C_\text{SP}  ,
   \end{align}
where $\rho$ is the liquid density, $\mu$ the viscosity, and $p$ the pressure. On the surface of the bubble:
   \begin{align}
\frac{\partial \Gamma_\text{SP} }{\partial t} + \nabla_s \cdot (\vec{u}_s \Gamma_\text{SP})=& D_s \nabla_s^2 \Gamma_\text{SP} + (\Gamma_m - \Gamma_T) \;k_{\text{ads}}^\text{SP} \; C_\text{SP} \nonumber \\ & - k_{\text{des}}^\text{SP} \; \Gamma_\text{SP} \; \exp\left(-\frac{A_\text{SP} \Gamma_T}{\Gamma_m}\right) 
+k_\text{fwd} \Gamma_\text{MCH} -k_\text{rev} \Gamma_\text{SP}  \\
\frac{\partial \Gamma_\text{MCH} }{\partial t} + \nabla_s \cdot (\vec{u}_s \Gamma_\text{MCH})&= D_s \nabla_s^2 \Gamma_\text{MCH} + (\Gamma_m - \Gamma_T) \;k_{\text{ads}}^\text{MCH} \; C_\text{MCH} \nonumber \\ & - k_{\text{des}}^\text{MCH} \; \Gamma_\text{MCH} \; \exp\left(-\frac{A_\text{MCH} \Gamma_T}{\Gamma_m}\right) 
-k_\text{fwd} \Gamma_\text{MCH} +k_\text{rev} \Gamma_\text{SP}  \\
\gamma = &\gamma_0 + nRT \; \ln \left(1 - \frac{\Gamma}{\Gamma_m}\right).
   \end{align}
The boundary conditions are, on the bubble surface:
\begin{align}
  -D \frac{\partial C_\text{SP} }{\partial n} &= (\Gamma_m - \Gamma_T) \;k_{\text{ads}}^\text{SP} \; C_\text{SP} - k_{\text{des}}^\text{SP} \; \Gamma_\text{SP} \; \exp\left(-\frac{A_\text{SP} \Gamma_T}{\Gamma_m}\right) \\
    -D \frac{\partial C_\text{MCH} }{\partial n} &= (\Gamma_m - \Gamma_T) \;k_{\text{ads}}^\text{MCH} \; C_\text{MCH}  - k_{\text{des}}^\text{MCH} \; \Gamma_\text{MCH} \; \exp\left(-\frac{A_\text{MCH} \Gamma_T}{\Gamma_m}\right) \\
    \vec{u} \cdot \vec{n} &= 0,
   \end{align}
where $\vec{n}$ denotes the direction normal to the surface and into the liquid. The (no-flux) boundary conditions on the solid surfaces are
\begin{align}
 \frac{\partial C_\text{SP} }{\partial n} &= 0 \\
     \frac{\partial C_\text{MCH} }{\partial n} &= 0\\
    \vec{u}  &= \vec{0}.
   \end{align}

\subsection{Reference Frames}
\begin{figure*}[ht]
\centering
\includegraphics[width=0.5\linewidth]{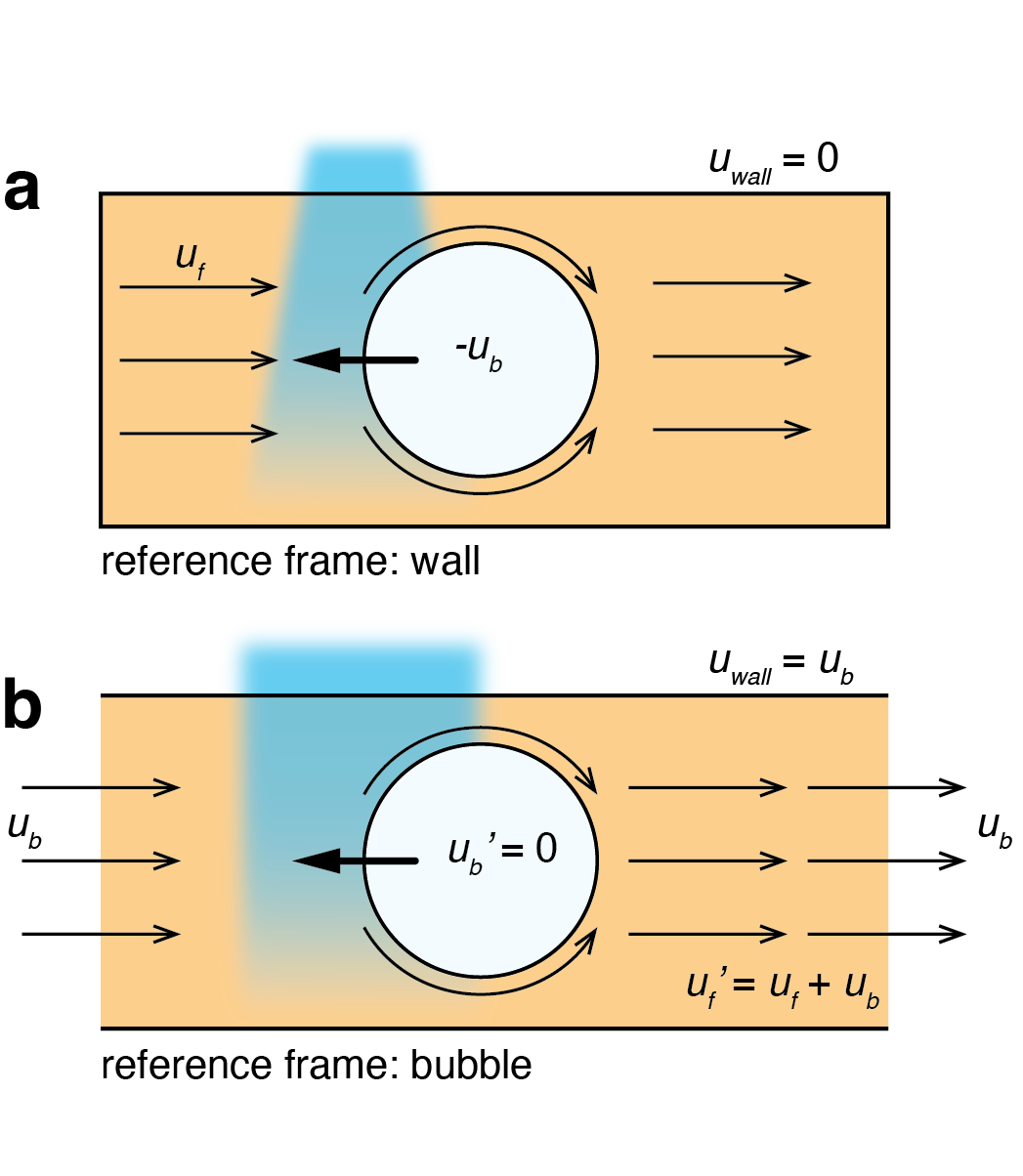}
\caption{\justifying
Reference frames in the experimental setup and modeling framework.
\textbf{a.} In the space experiment, a bubble is confined in a closed-ended cuvette with fixed walls ($u_{\text{wall}} = 0$). The bubble moves leftward at $-u_b$, while the fluid flows with velocity $u_f$.
\textbf{b.} In the numerical model, a steady-state, open-ended, two-dimensional model is constructed using the bubble as the reference frame. In this frame, the adjusted fluid velocity is $u_f' = u_f + u_b$, and the walls move at $u_{\text{wall}} = u_b$. The original fluid velocity in the experimental frame can be obtained by reversing this transformation.
}
\label{COMSOL: reference frame}
\end{figure*}

The reference frames in both the experimental setup and the modeling framework are illustrated in Figure \ref{COMSOL: reference frame}. In the space experiment (Figure \ref{COMSOL: reference frame}a), a bubble is sealed inside a closed-ended cuvette filled with test fluid. The walls are fixed (\( u_{\text{wall}} = 0 \)), and the bubble migrates to the left with a velocity of \( -u_b \) (the negative sign indicates leftward motion), while the fluid has a velocity \( u_f \). To simplify the model and reduce computational time, we construct a two-dimensional numerical model of steady-state with open ends, using the bubble as the reference frame (Figure \ref{COMSOL: reference frame}b). Light is moving with the bubble as it migrates. In this frame, the computed fluid velocity becomes the original fluid velocity minus the bubble's migration velocity (\( u_f' = u_f - (-u_b) = u_f + u_b \)), and the walls now move with a velocity of \( u_{\text{wall}} = 0 - (-u_b) = u_b \). The bubble velocity \( u_b \) can be determined by extracting the bulk fluid velocity at the inlet, and the fluid velocity in the original frame is obtained by reversing the transformation (\( u_f = u_f' - u_b \)). 

\captionsetup{justification=raggedright, singlelinecheck=false} 

\newpage
\section{Scaling theory for bubbles in liquid}

\textbf{Drag force} \\
To simplify the model and reduce computational time, the COMSOL simulation was conducted in 2D. As a result, a cylindrical geometry was used to account for the drag force in the system. 

The drag per unit length of a cylinder at low Reynolds number is expressed as \cite{huner1977cylinder}:  
\begin{equation}
D_{drag_L} = 4 \pi \mu U \varepsilon
\end{equation}
where $\mu$ is the dynamic viscosity of the fluid, $U$ is the flow velocity, and $\varepsilon$ is a factor defined as:
\begin{equation}
\varepsilon = \left[ \frac{1}{2} - \gamma - \ln\left(\frac{\text{Re}}{8}\right) \right]^{-1}
\end{equation}
with $\gamma$ representing Euler's constant, which is 0.577216, and $\text{Re}$ being the Reynolds number. 

The total drag force is then approximated as:
\begin{equation}
D_{drag} \sim \mu U \varepsilon L_z
\end{equation}
where $L_z$ is the characteristic length of the system.

\textbf{Propulsive Force}\\
The propulsive force exerted by the fluid is determined by the product of the shear stress and the area over which it acts:
\begin{equation}
F_{\text{propelling}} = \tau_{\text{prop}} \cdot A_{\text{prop}}
\end{equation}
where $\tau_{\text{prop}}$ is the shear stress, and $A_{\text{prop}}$ is the area in contact with the fluid.

\begin{figure*}[h]
\centering
\includegraphics[width=0.5\linewidth]{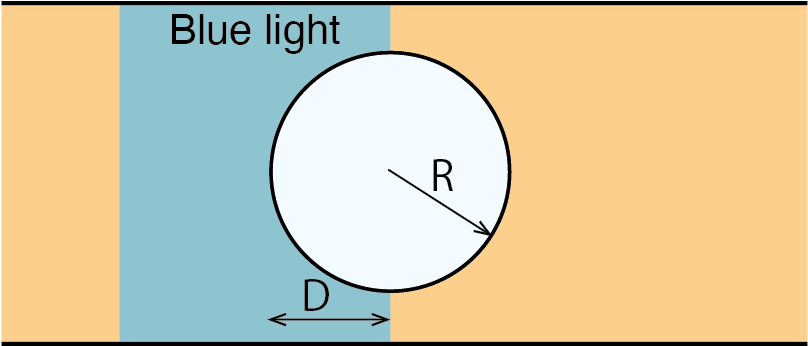}
\caption{\justifying
Scaling sketch showing the bubble radius $R$ and and the illuminated length $D$.
}
\label{fig:scaling sketch}
\end{figure*}

The area is approximately on the order of $(2R - D) L_z$, where $R$ is the bubble radius, and $D$ is the illuminated length (Figure \ref{fig:scaling sketch}).

The shear stress is estimated as:  
\begin{equation}
\tau_{\text{prop}} \sim \frac{\gamma_2 - \gamma_1}{2R - D},
\end{equation}
where $\gamma_1$ and $\gamma_2$ represent the surface tensions before and after illumination, respectively. Here, $\gamma_1$ represents the surface tension where the molecules have fully switched and just exited the illuminated zone. Thus, $\gamma_1$ corresponds to the minimum surface tension at a given concentration, which equals $\gamma_B$. 

In contrast, $\gamma_2$ is the surface tension at the rear point of the bubble, where the molecules are partially reverting towards their dark-state value. At this point, the surface tension lies between the values corresponding to the fully illuminated and dark states and is given by:

\begin{equation}
\gamma_2 = \gamma_A + (\gamma_B - \gamma_A) e^{-\frac{t_{12}}{\tau_{\text{rev}}} \cdot C_1}
\end{equation}

where $t_{12}$ is the time required for the molecules to travel from point 2 to point 1, approximated as $C_3\cdot(2R - D) / U$. $\tau_{\text{rev}}$ is the reverse time constant for surface tension recovery, and $C_1$ is a prefactor.

By multiplying $\tau_{\text{prop}}$ with $A_{\text{prop}}$ and substituting the expressions for $\gamma_1$ and $\gamma_2$, the propulsive force becomes:

\begin{equation}
F_{\text{propelling}} \sim \left[(\gamma_A - \gamma_B) \left(1 - e^{-\frac{t_{12}}{\tau_{\text{rev}}} \cdot C_1}\right)\right] \cdot L_z,
\end{equation}

The fluid resistance experienced by a self-propelled bubble moving through a liquid is balanced by the propelling force. This balance, expressed as $F_{\text{drag}} \sim F_{\text{Ma}}$, yields:

\begin{equation}
C_2 \mu U \varepsilon L_z \sim \left[(\gamma_A - \gamma_B) \left(1 - e^{-\frac{t_{12}}{\tau_{\text{rev}}} \cdot C_1}\right)\right] L_z,
\end{equation}

where $C_2$ is a prefactor. 

By substituting $t_{12} \sim \frac{2R - D}{U}$ into the equation, the product $C_1 \cdot C_3$ simplifies to a new prefactor $C_4$. The equation for the bubble velocity at low Re then becomes:

\begin{equation}
U = \frac{\gamma_A - \gamma_B}{\mu} \cdot \frac{1}{C_2 \cdot \varepsilon} \left(1 - e^{-C_4 \cdot \frac{2R - D}{U \cdot \tau_{\text{rev}}}}\right).
\end{equation}

This expression demonstrates how the bubble velocity depends on the bubble size, given a specific surface tension difference and fluid viscosity.

The model predicts the velocity profile over a range of bubble radii and compares it with experimental data. The velocity equation is solved iteratively by calculating the Reynolds number and applying a correction factor $\varepsilon$ to account for low-Reynolds-number flow. The parameters $C_2$ (associated with drag) and $C_4$ (related to Marangoni force decay) are initially guessed and refined to minimize the total error between the predicted and experimental velocities, with $C_2$ is 1058.6 and $C_4$ is 1.293.

\textbf{Reverse time constant} \\
This model predicts the time constant for the reverse reaction surface tension during the transient state using the Frumkin isotherm. The interconversion dynamics between MCH and SP were modeled with first-order kinetic equations. As noted earlier, concentration changes of $C_{\text{MCH}}$ and $C_{\text{SP}}$ are governed by:

\[
\frac{dC_{\text{MCH}}}{dt} = k_{\text{rev}} C_{\text{SP}} - k_{\text{fwd}} C_{\text{MCH}}, \quad \frac{dC_{\text{SP}}}{dt} = -\frac{dC_{\text{MCH}}}{dt},
\]

where $k_{\text{fwd}}$ and $k_{\text{rev}}$ are the forward and reverse reaction rate constants. The initial concentrations, $C_{\text{MCH},0}$ and $C_{\text{SP},0}$, depend on the switching factor, which determines the fraction of MCH converted to SP under illumination.

The surface coverage, $\theta(t)$, is modeled using the Frumkin isotherm to account for competitive adsorption and molecular interactions:

\begin{equation}
\theta = \frac{\kappa_F^A C_A + \kappa_F^B C_B}{\kappa_F^A C_A + \kappa_F^B C_B + e^{-\tilde{A} \theta}},
\label{eq:surface_kinetics_ss_theta_light}
\end{equation}

where $\kappa_F^A$ and $\kappa_F^B$ are adsorption coefficients for MCH and SP, and $\tilde{A}$ is a parameter reflecting intermolecular interactions. 

The surface tension, $\gamma(t)$, is predicted based on $\theta(t)$ using:

\begin{equation}
\gamma(t) = \gamma_0 + nRT \, \Gamma_m \left[ \ln (1 - \theta) - \tilde{A} \frac{\theta^2}{2} \right],
\label{eq:Frumkin_isotherm}
\end{equation}

where $\gamma_0$ is the initial surface tension, $nRT$ is the product of the stoichiometric factor $n$, the universal gas constant $R$, and temperature $T$, and $\Gamma_m$ is the maximum surface concentration. This equation captures the effects of unoccupied surface area and molecular interactions on surface tension.

The system of differential equations was integrated using MATLAB’s \texttt{ode45} solver to track the evolution of $C_{\text{MCH}}(t)$, $C_{\text{SP}}(t)$, $\theta(t)$, and $\gamma(t)$. These predictions provide insight into the transient surface tension behavior of light-switchable surfactants. The time constant for the reverse reaction was estimated to be 100 seconds.

\newpage
\section{Fluid velocity}

\captionsetup{justification=raggedright, singlelinecheck=false} 
\begin{figure*}[h!]
\centering
\includegraphics[width=\linewidth]{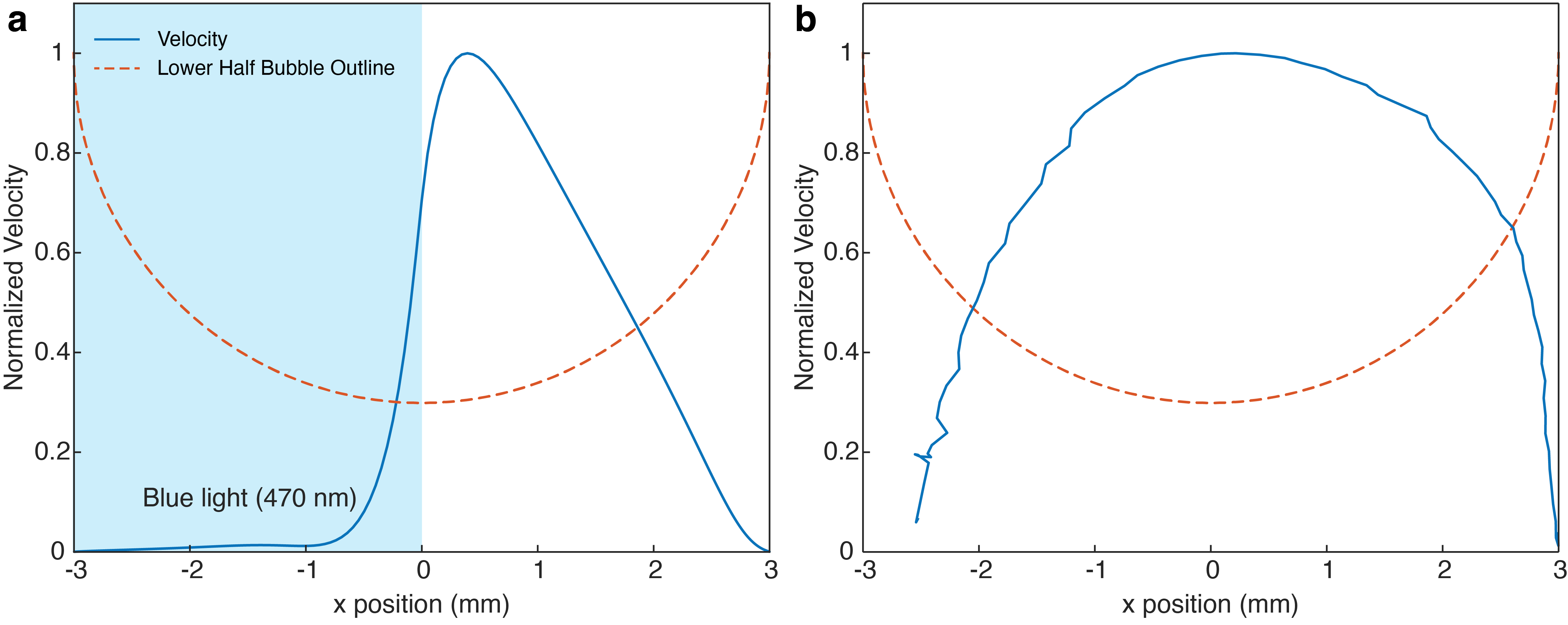}
\caption{\justifying
\textbf{a.} Simulated fluid velocity along the surface of a 3 mm radius bubble in a $1 \times 10^{-6}$ mM solution using COMSOL. 
\textbf{b.} Experimentally measured fluid velocity along the bubble surface from tracer particles. 
}
\label{fig:fluid velocity}
\end{figure*}

The fluid velocity along the bubble surface was simulated using COMSOL (Figure \ref{fig:fluid velocity}a). For a 3 mm radius bubble in a solution concentration of $1 \times 10^{-6}$ mM, a similar trend in fluid velocity along the surface was observed, as shown in Figure \ref{fig:fluid velocity}b. When the bubble is partially illuminated, the fluid velocity in the light-exposed bulk region is uniformly lower than in the transition zone between dark and light regions. This is due to the even distribution of switched SP molecules in the illuminated area. In the transition zone, the velocity increases due to a larger surface tension gradient, then decreases as molecules revert to the MCH form. Unlike the experimental fluid velocity obtained from tracer particles, the simulated bubble remains fixed, with the light position constant relative to the bubble. Consequently, the fluid speed continues to decrease with increasing distance from the light source.

\clearpage
\section{Supplementary Figures and Tables}

\begin{figure*}[h!]
\centering
\includegraphics[width=\linewidth]{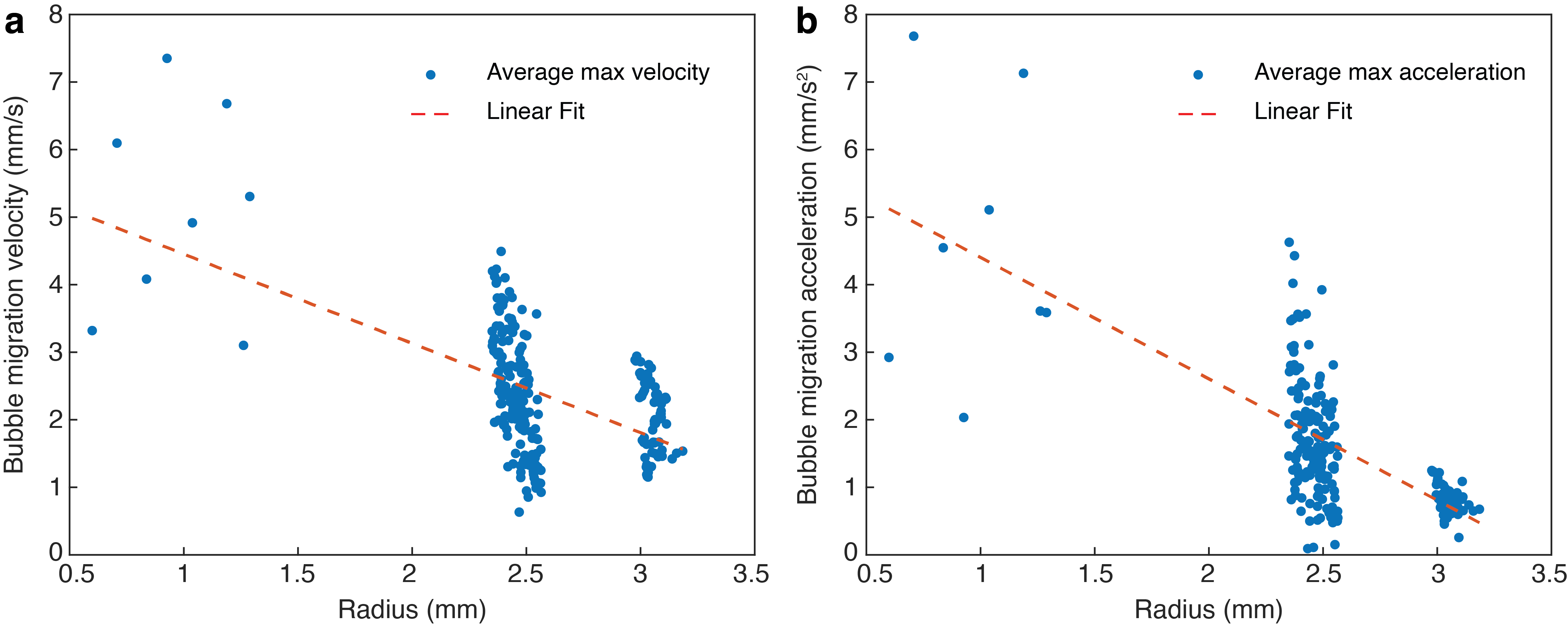}
\caption{
\textbf{a.} Average maximum migration velocity as a function of bubble radius. 
\textbf{b.} Average maximum acceleration as a function of bubble radius.
}
\label{fig: size effect(data points)}
\end{figure*}

\begin{figure*}[h!]
\centering
\includegraphics[width=0.5\linewidth]{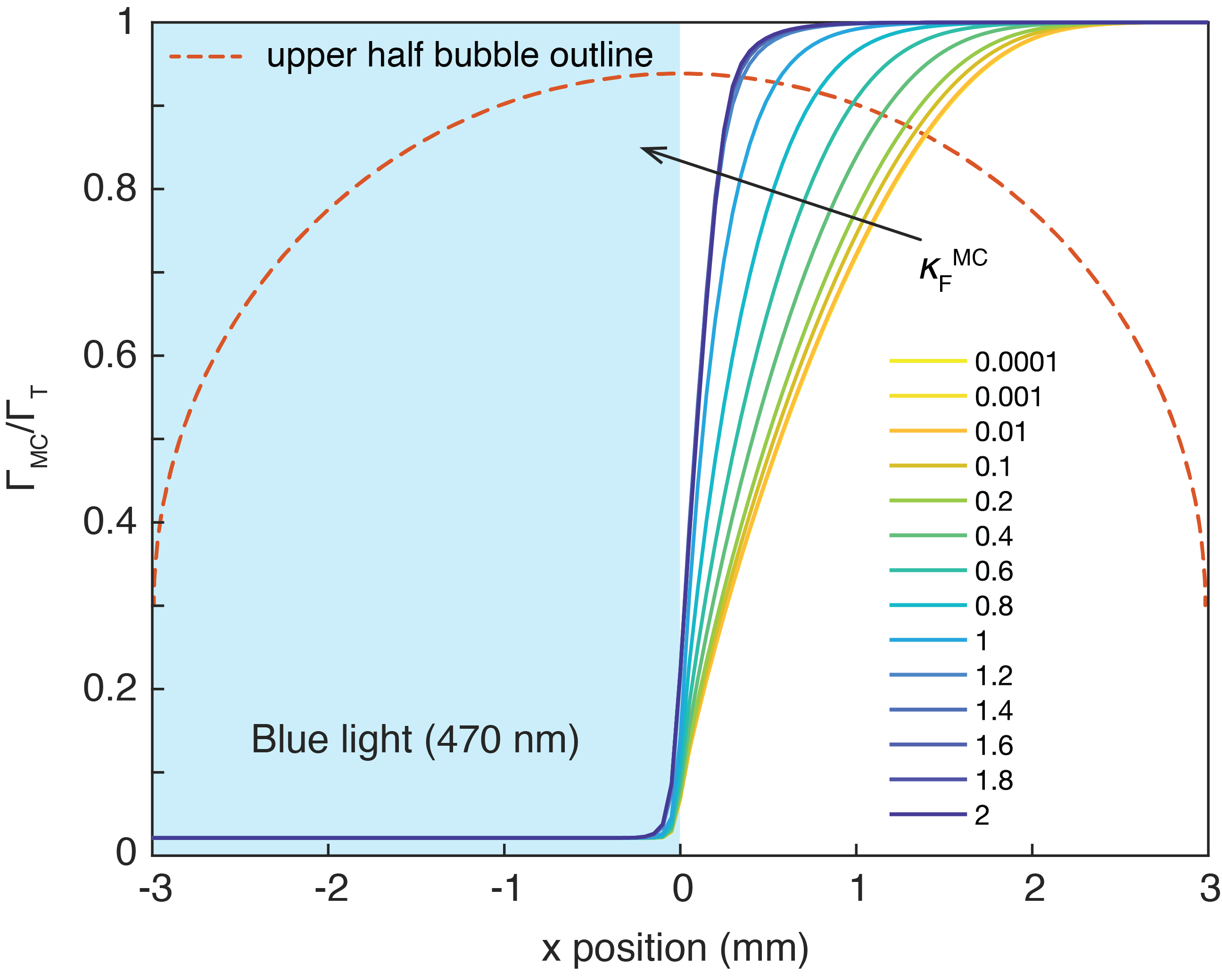}
\caption{Surface coverage of MCH-para (\(\Gamma_{\text{MCH}} / \Gamma_T\)) along the x-axis position of a bubble exposed to 470 nm blue light. The dashed line represents the outline of the upper half of the bubble. Curves correspond to different values of the adsorption constant \(\kappa_F^{\text{MCH}}\), illustrating variations in surface coverage.
}
\label{fig:surface coverage}
\end{figure*}

\clearpage

\section{Other Supporting Materials}

Movie S1: Migration of an air bubble in methanol containing 1 mM MCH-para.

\end{document}